\newif\ifconfver
\confverfalse      %declaring conference version false
\newif\ifcutshort      %this level shortens the equations
\cutshorttrue

\newif\ifcutshortlvltwo  %this level takes out some examples, figs., and sim.
\cutshortlvltwofalse

\ifconfver
    \documentclass[10pt,twocolumn,twoside]{IEEEtran}
\else
    \documentclass[11pt,draftcls,onecolumn]{IEEEtran}
\fi

\usepackage{cite}
\usepackage{graphicx}
\usepackage{psfrag}
\usepackage{subfigure}
\usepackage{url}
\usepackage{stfloats}
\usepackage{amsfonts,amssymb,amsmath,bm,paralist,theorem,cite,ifthen,color}
\usepackage{array}
\usepackage{caption}
\usepackage{calc}
\usepackage{subfig}
\usepackage{longtable}
\usepackage{stfloats}  % Written by Sigitas Tolusis
\usepackage{graphics,color,epsfig,subfigure,enumerate}
\usepackage{multirow}
\usepackage[algoruled,linesnumbered]{algorithm2e}

\newtheorem{Lemma}{Lemma}

{\begin{list}{}{
    \settowidth{\labelwidth}{\mbox{\textnormal{#1}}}%
    \setlength{\leftmargin}{\labelwidth+\labelsep}}}%
{\end{list}}

\newcommand\Cplx{\ensuremath{{\mathbb{C}}}}
\newcommand\vecc{\ensuremath{{\rm vec}}}
\newcommand\ab{\ensuremath{{\bm a}}}
\newcommand\Ab{\ensuremath{{\bm A}}}

\newcommand\eb{\ensuremath{{\bm e}}}
\newcommand\hb{\ensuremath{{\bm h}}}
\newcommand\rb{\ensuremath{{\bm r}}}
\newcommand\tb{\ensuremath{{\bm t}}}
\newcommand\wb{\ensuremath{{\bm w}}}

\newcommand\ub{\ensuremath{{\bm u}}}
\newcommand\xb{\ensuremath{{\bm x}}}

\newcommand\Cb{\ensuremath{{\bm C}}}
\newcommand\Wb{\ensuremath{{\bm W}}}
\newcommand\Qb{\ensuremath{{\bm Q}}}
\newcommand\Ub{\ensuremath{{\bm U}}}
\newcommand\Lambdab{\ensuremath{{\bm \Lambda}}}
\newcommand\Prob{\ensuremath{{\rm Prob}}}
\newcommand\zerob{\ensuremath{{\bm 0}}}
\newcommand\Ib{\ensuremath{{\bm I}}}

\newcommand\Ral{\ensuremath{{\mathbb{R}}}}
\newcommand\tr{\ensuremath{{\rm Tr}}}
\newcommand\Diag{\ensuremath{{\rm Diag}}}

\begin{document}

\bibliographystyle{IEEEtran}
% paper title

\title{Outage Constrained Robust Transmit Optimization for Multiuser MISO Downlinks: Tractable Approximations by Conic Optimization}

\ifconfver \else {\linespread{1.1} \rm \fi

\author{\vspace{0.5cm} Kun-Yu Wang, Anthony Man-Cho So, Tsung-Hui Chang, Wing-Kin
Ma, and Chong-Yung Chi

\thanks{$^\S$This work is supported
by the National Science Council, R.O.C.,
under Grant NSC-99-2221-E-007-052-MY3; and
partly by The Chinese University of Hong Kong, under Direct Grant
2050396; and by the Shun Hing Institute of Advanced Engineering at The Chinese University of Hong Kong, under Project \#MMT-p2-09.
Part of this work has been published in EUSIPCO
2010 \cite{Wang2010} and in IEEE ICASSP 2011 \cite{Wang2011}.}
\thanks{K.-Y. Wang, T.-H. Chang and C.-Y. Chi are with the
Institute of Communications Engineering \& Department of Electrical
Engineering, National Tsing Hua University, Hsinchu, Taiwan 30013,
R.O.C. E-mail:
kunyuwang7@gmail.com,~tsunghui.chang@gmail.com,~cychi@ee.nthu.edu.tw.}

\thanks{A. M.-C. So is with the Department of Systems Engineering
and Engineering Management and the Shun Hing Institute of Advanced
Engineering, The Chinese University of Hong Kong, Hong Kong. E-mail:
manchoso@se.cuhk.edu.hk.}

\thanks{W.-K. Ma is the corresponding author. Address:
Department of Electronic Engineering, The Chinese University of Hong
Kong, Shatin, Hong Kong S.A.R., China. E-mail: wkma@ieee.org.}
 }

\maketitle

\vspace{-0.3cm}
\begin{abstract}
In this paper we consider a probabilistic signal-to-interference-and-noise ratio (SINR)
constrained problem for transmit beamforming design
in the presence of imperfect channel state information (CSI),
under a multiuser multiple-input single-output (MISO) downlink scenario.
In particular, we deal with outage-based quality-of-service constraints,
where the probability of each user's SINR not satisfying a service requirement must not fall below a given outage probability specification.
The study of solution approaches to the probabilistic SINR constrained problem is important because CSI errors are often present in practical systems and they may cause substantial SINR outages if not handled properly.
However, a major technical challenge is how to process the probabilistic SINR constraints.
To tackle this, we propose a novel relaxation-restriction (RAR)
approach, which consists of two key ingredients---
semidefinite relaxation (SDR),
and analytic tools for conservatively approximating probabilistic constraints.
The underlying goal is to establish approximate probabilistic SINR constrained formulations in the form of convex conic optimization problems, so that they can be readily implemented by available solvers.
Using either an intuitive worst-case argument or specialized probabilistic results, we develop various conservative approximation schemes for processing probabilistic constraints with quadratic uncertainties.  Consequently, we obtain several RAR alternatives for handling the probabilistic SINR constrained problem.
Our techniques apply to both complex Gaussian CSI errors and i.i.d. bounded CSI errors with unknown distribution.
%Although the development mainly focuses on a complex Gaussian CSI error model,
%an interesting case study of i.i.d. bounded CSI errors with unknown distribution is also provided.
Moreover, results obtained from our extensive simulations show that the proposed RAR methods significantly improve upon existing ones, both in terms of solution quality and computational complexity.%, when applied to the probabilistic SINR constrained problem.

\end{abstract}

\ifconfver \else \IEEEpeerreviewmaketitle} \fi

%---------------------------------------------------------------------------
\ifconfver \else
\newpage
\fi

\vspace{-0.0cm}
%%%%%%%%%%%%%%%%%%%%%%%%%%%%%%%%%%%%%%%%%%%%%%%%%%%%%%%%%%%%%%%%%%
\section{Introduction}
%%%%%%%%%%%%%%%%%%%%%%%%%%%%%%%%%%%%%%%%%%%%%%%%%%%%%%%%%%%%%%%%%%

In multi-antenna multiuser downlinks,
linear transmit beamforming has been recognized as an important technique,
capable of leveraging quality of service (QoS)
and increasing limits on the number of users served;
see, e.g., the review article \cite{Gershman2010_SPM}
and the references therein.
Transmit beamforming design approaches developed in this context
have not only been proven to offer efficient and flexible solutions for
QoS optimization and interference management in standard downlinks, but
have also been modified or generalized to deal with designs arising from frontier scenarios, such as relay networks~\cite{Gershman2010_SPM}, cognitive radios~\cite{ZLC10}, and multicell coordinated downlinks~\cite{BK:Bengtsson01,DahroujY10TWC,GHH+10}.

In transmit beamforming,
a very representative problem setting is the unicast multi-input single-output (MISO) downlink scenario,
wherein a multi-antenna base station simultaneously transmits data streams to a number of single-antenna users, each stream for a designated user, by carefully directing transmit beams to the users.
The problem of interest is to provide a signal-to-interference-and-noise ratio (SINR) constrained design formulation,
in which transmit beamformers for the users are sought, so that
each user is served with a QoS, characterized by the SINR, no less than a prescribed requirement,
and that the transmit power is minimized.
The SINR constrained problem is a meaningful and frequently used design formulation in practice,
and essentially the same problem formulation can be seen in other works,
such as those in the aforementioned frontier scenarios~\cite{DahroujY10TWC,Zheng_etal2009,Chalise09TSP}.
It is also a fundamentally intriguing problem.
There are three parallel solution approaches to the problem,
namely, uplink-downlink duality~\cite{Farrokhi1998,Schubert2004},
semidefinite relaxation (SDR)~\cite{BK:Bengtsson01,Huang10TSP}, and the second-order cone programming (SOCP) formulation~\cite{WieselTSP06}.
Each of those approaches is elegant, offering different implications both in theory and in practical implementations.
They also serve as stepping stones for more advanced designs,
such as those under imperfect channel state information (CSI) effects.

The SINR constrained problem, like many other transmit optimization problems,
is based on the assumption that the downlink CSI is perfectly available at the base station.
Unfortunately, such an assumption generally does not hold in practice~\cite{Love2008}.
In the time division duplex (TDD) setting, where there is a reciprocity between the uplink and downlink channels, the downlink CSI is typically acquired by uplink channel estimation from training data.  Channel estimation errors, which are caused by noise and a limited amount of training data, result in CSI errors in this setting.
In the frequency division duplex (FDD) setting,
CSI acquisition is often achieved by CSI feedback with limited rates.  As a result, quantization errors arising from the limited feedback lead to imperfect CSI.
In addition, CSI may become somewhat outdated if the user mobility speed happens to be faster than the CSI update speed.
If one uses the corrupt CSI directly to design the transmit beamformers, then the users may experience severe SINR outages and not be able to receive their anticipated QoS levels.

Recently, there has been much attention on transmit beamforming designs that are robust against CSI errors.
In particular, it is of significant interest to consider ``safe'' SINR constrained formulations under various CSI error models,
where users' SINR requirements must be satisfied
even with the worst possible CSI errors, or, alternatively, with high probability.
%Such robust formulations depend on the CSI error model.
One commonly considered formulation at present is the worst-case SINR constrained problem, in which the CSI errors are assumed to lie in a bounded set (known as the \emph{uncertainty set}).
This worst-case robust problem appears to be a hard (nonconvex) problem, since the worst-case SINR constraints are semi-infinite and indefinite quadratic.
Several concurrent approximation schemes have been proposed to tackle the worst-case robust problem;
notable works include the conservative SOCP formulation~\cite{Shenouda2007}, the robust MMSE formulation~\cite{Boche2009,Shenouda2009_WorstcaseMSE}, and SDR~\cite{ZhengWongNg_2008}.
The beauty of these works lies in the careful combination of robust optimization results~\cite{BEN09} and problem formulations, leading to convex and tractable design solutions.

Another safe formulation, which is the focus of this paper,
is the {\it probabilistic}, or {\it outage-based, SINR constrained problem}.
In this formulation, we assume a random CSI error model,
such as the popular complex Gaussian model,
and the SINR outage probability of each user must be kept below a given specification.
Unfortunately, while the worst-case SINR constrained problem is considered hard to solve already,
this is even more so with the probabilistic SINR constrained problem---
Probabilistic SINR constraints generally have no closed form expressions
and are unlikely to be easily handled in an exact way.
Thus, one has to resort to approximate design solutions.
To date, there are very few works on the probabilistic SINR constrained problem under the unicast downlink scenario.
In \cite{Vucic2009_chanceMSE}, the authors fix the transmit beam directions as zero forcing and then deal with a probabilistic power control problem.
In \cite{Shenouda2008}, a conservative SOCP formulation is developed using some advanced results in chance constrained optimization~\cite{Ben-tal2009,S09}.  A similar approach is presented in~\cite{Shenouda2009}, where the robust MMSE formulation is considered.

In this paper we propose several convex optimization solutions
for approximating the probabilistic SINR constrained problem.
Our approach is based on a relaxation-restriction (RAR) methodology.
Specifically, in the relaxation step, we employ SDR to linearize the quadratic terms in the SINR expression.  However, this step alone does not lead to an efficiently solvable formulation, because the probabilistic constraints imposed on the linearized SINR expressions are still nonconvex.
We circumvent this problem in the restriction step, where we first derive various analytic upper bounds on the {\it violation probability} (i.e., the probability that the constraints on the linearized SINR expressions are violated).  Such upper bounds serve as sufficient conditions for the probabilistic constraints to hold, hence the term ``restriction''.  Next, we show that our derived bounds are efficiently computable, which, together with the results from the relaxation step, leads to efficiently solvable approximations of the original probabilistic SINR constrained beamforming problem.  It should be noted that the above restriction approach has many advantages.  First, it allows one to generate feasible solutions to the probabilistic constraints, even when there is no closed form expression for the violation probability, or when the closed form expression is not efficiently computable.  Secondly, while it may be difficult to derive closed form expressions for the violation probability, it is usually much easier to derive upper bounds on it, thanks to the many powerful techniques from the probability theory literature.  Thirdly, there is usually more than one way to derive upper bounds on the violation probability, and this offers the possibility of trading approximation performance with computational complexity.  These advantages will become clear in our subsequent exposition.

The rest of this paper is organized as follows. The problem statement of the outage-based SINR constrained robust beamforming design problem is given in Section
\ref{sec: problem statement}. The idea of the proposed RAR method is introduced in Section \ref{sec: RAR}. In Sections \ref{sec: sphere bounding} and \ref{sec: probability inequality approaches}, various RAR formulations for complex Gaussian CSI errors are developed using either robust optimization or probabilistic techniques.  An RAR formulation for i.i.d.~bounded CSI errors is also presented in Section \ref{sec: probability inequality approaches}.  Simulation results are then presented in Section \ref{sec: simulation results}, and conclusions are drawn in Section \ref{sec: conclusions}.

%the sphere bounding-based RAR formulation is proposed. In Section , the Bernstein-type inequality-based and the decomposition-based RAR formulations are presented. The RAR formulation for i.i.d. bounded CSI errors is also presented in that section.

\emph{Notations}: We use boldfaced lowercase letters, e.g., $\ab$,
to represent vectors and uppercase letters, e.g., $\Ab$, to
represent matrices. The notations $\mathbb{R}^n$,
$\Cplx^n$, $\mathbb{S}^{n}$, and $\mathbb{H}^{n}$ stand for the sets
of $n$-dimensional real vectors, complex vectors, real symmetric
matrices and complex Hermitian matrices, respectively. The
superscripts `$T$' and `$H$' represent the transpose and (Hermitian)
conjugate transpose, respectively. $\Ab\succeq \zerob$ means that the
matrix $\Ab$ is positive semidefinite. $\tr(\Ab)$ and
$\lambda_{\max}(\Ab)$ denote the trace and maximum eigenvalue of
$\Ab$, respectively. $\vecc(\Ab)$ stands for the vector obtained by stacking
the column vectors of $\Ab$. $[\ab]_{i}$ and $[\Ab]_{ij}$ (or simply $a_i$ and $A_{ij}$) stand for
the $i$th entry of $\ab$ and $(i,j)$th entry of $\Ab$, respectively.
For a complex $\Ab$, we denote by ${\rm Re}\{\Ab\}$ and ${\rm
Im}\{\Ab\}$ its real and imaginary parts, respectively.
$\Ib_n$ denotes the $n \times n$ identity matrix.
Given scalars $a_1,\ldots,a_n$, we use $\Diag(a_1,\ldots,a_n)$ to
denote the $n\times n$ diagonal matrix whose $i$th diagonal entry is $a_i$.
$\|\cdot\|$ and
$\|\cdot\|_{F}$ represent the vector Euclidean norm and matrix
Frobenius norm, respectively. ${\mathbb E}\{\cdot\}$,
$\Prob\{\cdot\}$, and $\exp(\cdot)$ denote the statistical
expectation, probability function and exponential function,
respectively. We write $\xb\sim\mathcal{CN}(\bm{\mu},\Cb)$ if $\xb - \bm{\mu}$
is a circular symmetric complex Gaussian random vector with
covariance matrix $\Cb\succeq \zerob$.

\section{Problem Formulation}\label{sec: problem statement}
%%%%%%%%%%%%%%%%%%%%%%%%%%%%%%%%%%%%%%%%%%%%%%%%%%%%%%%%%%%%%%%%%%

We focus on a downlink multiuser MISO scenario, in which the base
station, or the transmitter, sends parallel data streams to multiple
users over the same channel. The transmission is unicast; i.e., each
data stream is exclusively for one user. The base station is
equipped with $N_t$ transmit antennae and the signaling strategy is
transmit beamforming. Let $\xb(t) \in \mathbb{C}^{N_t}$
denote the multi-antenna transmit signal vector of the base station
at time $t$. We have the following transmit signal model:
\begin{equation} \label{eq:model_x}
\xb(t)= \sum_{k=1}^K \wb_k s_k(t),
\end{equation}
where $\wb_k \in \mathbb{C}^{N_t}$ is the transmit beamforming
vector for user $k$, $K$ is the number of users, and $s_k(t)$ is the
user-$k$ data stream, which is assumed to have zero mean and unit
power (i.e., $\mathbb{E}\{ | s_k(t) |^2 \} = 1$). It is also assumed
that $s_k(t)$ is statistically independent of one another. For user
$i$, the received signal can be modeled as
%The received signal of user $i$ is modeled as
\begin{equation} \label{eq:model_y}
y_i(t)= \hb_i^H \xb(t) + n_i(t),
\end{equation}
where $\hb_i \in \mathbb{C}^{N_t}$ is the channel from the
base station to user $i$, and $n_i(t)$ is an additive noise, which is
assumed to have zero mean and variance $\sigma_i^2 > 0$.

A common assumption in transmit beamforming is that the base station
has perfect knowledge of $\hb_1,\ldots,\hb_K$; i.e., the so-called
perfect CSI setting.
%Such knowledge would then be exploited to design the beamforming vectors $\wb_1,\ldots,\wb_K$.
However, as discussed in detail in the Introduction, the base
station may not have perfect CSI in general. In this work, the CSI
is modeled as follows:
%\begin{equation} \label{eq:h}
$$ \hb_i = \bar{\hb}_i + \eb_i, \qquad i=1,\ldots,K, $$
%\end{equation}
where $\hb_i \in \mathbb{C}^{N_t}$ is the actual channel,
$\bar{\hb}_i \in \mathbb{C}^{N_t}$ is the presumed channel at the
base station (also called the imperfect CSI), and $\eb_i \in
\mathbb{C}^{N_t}$ is the respective error that is assumed to be
random. Our development will concentrate mainly on complex Gaussian
CSI errors, which is a commonly adopted model. Specifically, we
assume that
%\begin{equation}
$$ \eb_i \sim \mathcal{CN}( \bm{0}, \Cb_i ) $$
%\end{equation}
for some known error covariance $\Cb_i \succeq \bm{0}$,
$i=1,\ldots,K$.

%{\color{red} Regarding the distribution of $\eb_i$, we will
%be interested in the {\it complex circular Gaussian model}; i.e,
%\begin{equation}
%\eb_i \sim \mathcal{CN}( \bm{0}, \Cb_i )
%\end{equation}
%for some known error covariance $\Cb_i \in \mathbb{H}_{++}^{N_t}$, for $i=1,\ldots,K$.
%Note that the complex Gaussian model is a popular CSI error distribution model.
%In the later part of the paper, specifically, Sections XX-XXX,
%we will also consider other CSI error distribution models,
%such as the elementwise i.i.d. uniform model, and even an i.i.d. model with unknown distribution.
%}

%Two kinds of distribution models will be considered: i) {\it a complex circular Gaussian model},
%where we have $\eb_i \sim \mathcal{CN}( \bm{0}, \Cb_i )$ for some known error covariance $\Cb_i \in \mathbb{H}_{++}^{N_t}$;
%and ii) {\it an elementwise i.i.d. uniform model}, where each entry of $\eb_i$ has its real and imaginary parts being i.i.d. uniformly distributed.
%{\color{red} (talk more about the practical validity of the model? I fear that the discussion would be too long)}.

The goal here is to design beamforming vectors $\wb_1,\ldots,\wb_K$ such that
the QoS of each user
satisfies a prescribed set of requirements under imperfect CSI,
while
using the least possible amount of transmit power in doing so.
%keeping the total transmit power to the minimum.}
%minimizing the total transmit power at the same time.}
%And we wish to do so with the least amount of transmit power.
To put this into context, let us consider users' SINRs.  Under the model in
\eqref{eq:model_x}-\eqref{eq:model_y} and the associated
assumptions, the SINR of user $i$ is
%\begin{equation}
$$ \textsf{SINR}_i = \frac{ | \hb^H_i \wb_i |^2 }{ \sum_{k \neq i} |
\hb^H_i \wb_k |^2 + \sigma_i^2 }. $$
%\end{equation}
To accommodate
imperfect CSI knowledge at the base station, which causes uncertainties in the
actual SINRs, we consider the following robust beamforming design
problem:

~ \\
\noindent \fbox{\parbox{\linewidth}{ {\bf Probabilistic SINR
constrained problem:} \ Given minimum SINR requirements
$\gamma_1,\ldots,\gamma_K > 0$ and maximum tolerable outage
probabilities $\rho_1,\ldots,\rho_K \in (0,1]$, solve
%\begin{subequations} \label{eq:main}
%\begin{align}
%\min_{ \wb_1,\ldots, \wb_K \in \mathbb{C}^{N_t} } & ~ \sum_{i=1}^K \| \wb_i \|^2 \\
%{\rm s.t.} & ~ {\rm Prob}\left\{  \frac{ | (\bar{\hb}_i + \eb_i)^H \wb_i |^2 }{ \sum_{k \neq i} | (\bar{\hb}_i + \eb_i)^H \wb_k |^2 + \sigma_i^2 } \geq \gamma_i  \right\} \geq 1 - \rho_i, \quad i=1,\ldots,K.
%\label{eq:main_SINR_con}
%\end{align}
%\end{subequations}
\begin{subequations} \label{eq:main}
\begin{align}
\min_{ \wb_1,\ldots, \wb_K \in \mathbb{C}^{N_t} } & ~ \sum_{i=1}^K \| \wb_i \|^2 \\
{\rm s.t.} & ~ {\rm Prob}_{ \hb_i \sim
\mathcal{CN}(\bar{\hb}_i,\Cb_i) } \left\{  {\sf SINR}_i \geq
\gamma_i  \right\} \geq 1 - \rho_i, \quad i=1,\ldots,K.
\label{eq:main_SINR_con}
\end{align}
\end{subequations}
}} ~

%Formulation \eqref{eq:main} aims to provide QoS in an outage
%probability limited sense--- for each user, say, user $i$, there is
%an SINR guarantee of no less than $\gamma_i$ in at least $(1-\rho_i)
%\times 100 \%$ of the time.
%%{\color{red} \it (We may need to shed
%%some light on motivations and related works. I plan to do it in the
%%introduction, and perhaps will have a little say right over here.
%%Anyway, I will do it after the introduction is done.) }
%{\color{blue}
%Physically speaking, a smaller $\rho_i$ would mean higher QoS fidelity.
%In fact, the simulation results in Section~\ref{sec: simulation results} will
%demonstrate that a ``non-robust design'';
%i.e., running a perfect-CSI SINR constrained problem where
%the actual channels ${\hb}_i$ are substituted by the presumed channels $\bar{\hb}_i$,
%can suffer from serious SINR outage.
%But it should be noted that setting $\rho_i$ too small may also make the design overly conservative, leading to a large transmit power requirement, or even no feasible solution, with problem~\eqref{eq:main}.
%Hence, there is a tradeoff between QoS fidelity and design conservativeness,
%as governed by the chosen outage probability specification $\rho_i$.
%}

Formulation~\eqref{eq:main} is an instance of the so-called {\it chance constrained optimization problem} due to the presence of the probabilistic constraints~\eqref{eq:main_SINR_con}, and it will be the main focus of this paper.
In~\eqref{eq:main},
the design parameters $\rho_i$'s
govern service fidelity,
making sure that each user, say, user $i$,
is served with an SINR no less than $\gamma_i$
at least $(1-\rho_i) \times 100 \%$ of the time.
%the QoS is defined in an outage probability limited sense---
%a feasible solution of \eqref{eq:main} guarantees that
%every user, say, user $i$, is served with an SINR no less than $\gamma_i$
%in at least $(1-\rho_i) \times 100 \%$ of the time.
%Hence, a smaller $\rho_i$ means a higher service fidelity.
In fact, the simulation results in Section~\ref{sec: simulation results} will
demonstrate that a ``non-robust design'';
i.e., designing the beamformers by running the perfect-CSI-based SINR constrained problem
with actual channels ${\hb}_i$ substituted by the presumed channels $\bar{\hb}_i$,
can suffer from serious SINR outage.
Moreover, it should be noted that there is a tradeoff between service fidelity and design conservatism.
On one hand, it is desirable to request higher service fidelity by using small values with $\rho_i$'s.
On the other hand, the design in \eqref{eq:main} would become more conservative as $\rho_i$'s decrease.
In particular, for very small $\rho_i$'s, one may end up with design solutions that have unacceptably large transmit power, or there may be no feasible solution to~\eqref{eq:main}.
%(even though it is optimal to \eqref{eq:main}),
%In particular, if $\rho_i$'s are set too small,
%then the design in \eqref{eq:main} would become overly conservative.

Although the probabilistic SINR constrained problem in
\eqref{eq:main} is a meaningful design criterion, it is a very hard
problem. The main difficulty is that the probability functions in
\eqref{eq:main_SINR_con} do not yield simple closed form expressions for
the considered CSI error distribution models.  Thus, one may only
resort to approximation methods. In the next section we will
describe our proposed approximation approach.

\vspace{-0.3cm}
%%%%%%%%%%%%%%%%%%%%%%%%%%%%%%%%%%%%%%%%%%%%%%%%%%%%%%%%%%%%%%%%%%
\section{The Relaxation-Restriction Approach}\label{sec: RAR}
%%%%%%%%%%%%%%%%%%%%%%%%%%%%%%%%%%%%%%%%%%%%%%%%%%%%%%%%%%%%%%%%%%

To handle the main problem~\eqref{eq:main}, we propose a novel relaxation-restriction (RAR) approach.
%Our proposed approach for handling the main problem \eqref{eq:main}
%is to employ a relaxation-restriction (RAR) rationale.
RAR-based
methods feature the use of convex optimization techniques to
approximate problem \eqref{eq:main}. Hence, they can be
efficiently implemented by available convex optimization software.

%%%%%%%%%%%%%%%%%%%%%%%%%%%%%%%%%%%%%%%%%%%%%%%%%%%%%%%%%%%%%%%%%%
\subsection{{Relaxation Step}}
%%%%%%%%%%%%%%%%%%%%%%%%%%%%%%%%%%%%%%%%%%%%%%%%%%%%%%%%%%%%%%%%%%

Let us first elaborate on the first step of RAR--- relaxation. The
motivation is that for each $i$, the inequality ${\sf SINR}_i \geq
\gamma_i$ is nonconvex in $\wb_1,\ldots,\wb_K$;
specifically, it is indefinite quadratic.
We handle this issue by
semidefinite relaxation (SDR)~\cite{Gershman2010_SPM,Luo2010_SPM}.
To illustrate SDR for the probabilistic SINR constrained problem,
we note that problem \eqref{eq:main} can be equivalently represented
by
\begin{subequations} \label{eq:main_alt}
\begin{align}
\min_{ \Wb_1,\ldots, \Wb_K \in \mathbb{H}^{N_t} } & ~ \sum_{i=1}^K {\rm Tr}( \Wb_i ) \\
{\rm s.t.} & ~ {\rm Prob}\left\{ ( \bar{\hb}_i + \eb_i )^H \left(
\frac{1}{\gamma_i} \Wb_i - \sum_{k \neq i} \Wb_k \right) (
\bar{\hb}_i + \eb_i ) \geq \sigma_i^2
\right\} \geq 1 - \rho_i, \quad i=1,\ldots,K, \\
& ~ \Wb_1, \ldots, \Wb_K \succeq \bm{0}, \\
& ~ {\rm rank}(\Wb_i)= 1, \quad i=1,\ldots,K,
\end{align}
\end{subequations}
where the connection between \eqref{eq:main} and \eqref{eq:main_alt} lies
in the feasible point equivalence
\[ \Wb_i = \wb_i \wb_i^H, \quad i=1,\ldots,K. \]
The SDR of \eqref{eq:main_alt} works by removing the nonconvex
rank-one constraints on $\Wb_i$; i.e., to consider the relaxed problem
\begin{subequations} \label{eq:sdr}
\begin{align}
\min_{ \Wb_1,\ldots, \Wb_K \in \mathbb{H}^{N_t} } & ~ \sum_{i=1}^K {\rm Tr}( \Wb_i ) \\
{\rm s.t.} & ~ {\rm Prob}\left\{ ( \bar{\hb}_i + \eb_i )^H \left(
\frac{1}{\gamma_i} \Wb_i - \sum_{k \neq i} \Wb_k \right) (
\bar{\hb}_i + \eb_i ) \geq \sigma_i^2 \right\} \geq 1 - \rho_i,
\quad i=1,\ldots,K,
\label{eq:sdr_SINR_con}  \\
& ~ \Wb_1, \ldots, \Wb_K \succeq \bm{0}.
\end{align}
\end{subequations}
The merit of this relaxation is that the inequalities inside the
probability functions in~\eqref{eq:sdr_SINR_con} are linear in
$\Wb_1,\ldots,\Wb_K$, which makes the probabilistic constraints in~\eqref{eq:sdr_SINR_con} more manageable. An issue that comes with SDR is the solution
rank--- the removal of ${\rm rank}(\Wb_i)=1$ means that the
solution $(\Wb_1,\ldots,\Wb_K)$ to problem \eqref{eq:sdr} may have rank higher than one.  We shall come back to this issue after presenting the restriction step.

%%%%%%%%%%%%%%%%%%%%%%%%%%%%%%%%%%%%%%%%%%%%%%%%%%%%%%%%%%%%%%%%%%
\subsection{{An Information Theoretic Interpretation of the Relaxation
Step}}
%%%%%%%%%%%%%%%%%%%%%%%%%%%%%%%%%%%%%%%%%%%%%%%%%%%%%%%%%%%%%%%%%%

{The SDR problem~\eqref{eq:sdr} has an alternative
interpretation from an information-theoretic point of view. Here, we
briefly describe this interpretation and the resulting implications
before moving to the restriction step of RAR.  Consider a general
transmission model:
\[ \xb(t)= \sum_{i=1}^K \xb_i(t), \]
where $\xb_i(t) \in \mathbb{C}^{N_t}$ is the transmit signal
intended for user $i$. Note that in contrast  to the original
transmission model in \eqref{eq:model_x}, where we fix the transmit
scheme as beamforming by setting $\xb_i(t) = \wb_i s_i(t)$, here we
do not assume any specific transmit structure.
%we do not assume a specific transmit structure,
%unlike the original transmission model in \eqref{eq:model_x}
%where we already fix the transmit scheme as beamforming by setting $\xb_i(t) = \wb_i s_i(t)$.
As a slight abuse of notations, let $\Wb_i = \mathbb{E}\{ \xb_i(t)
\xb_i^H(t) \}$ be the transmit covariance corresponding to user $i$.
From an information-theoretic perspective, the achievable rate of
user $i$ may be formulated as
\begin{align}\label{eq: rate}
  {\sf R}_i= \log_2\left(1+ \frac{\hb_i^H\Wb_i\hb_i}{\sum_{k\neq
  i}\hb_i^H\Wb_k\hb_i+\sigma_i^2}\right), %~\text{bits per channel   use.}
\end{align}
where the rates in \eqref{eq: rate}, in bits per channel use, are
achieved when $\xb_i(t)$ are Gaussian distributed (i.e., Gaussian
codebook). One can easily verify that the SDR problem~\eqref{eq:sdr} is precisely the following rate optimization problem:
\begin{subequations} \label{eq:it}
\begin{align}
\min_{ \Wb_1,\ldots, \Wb_K \in \mathbb{H}^{N_t} } & ~ \sum_{i=1}^K {\rm Tr}( \Wb_i ) \\
{\rm s.t.} & ~ {\rm Prob}_{ \hb \sim \mathcal{CN}( \bar{\hb}, \Cb_i
) } \left\{ {\sf R}_i \geq \log_2(1+ \gamma_i) \right\} \geq 1 -
\rho_i, \quad i=1,\ldots,K,
  \\
& ~ \Wb_1, \ldots, \Wb_K \succeq \bm{0}.
\end{align}
\end{subequations}
Specifically, the SDR equivalent \eqref{eq:it} is a total power
minimization problem that aims to ensure that each user is served
with a minimal rate of $\log_2(1+\gamma_i)$ bits per channel use,
with an outage probability no greater than $\rho_i$.

With this interpretation of the SDR, we can deduce an interesting implication:
If the SDR solution $(\Wb_1,\ldots,\Wb_K)$ to~\eqref{eq:sdr} does
not yield a rank-one structure, then an alternative to transmit
beamforming is to find another practical physical-layer scheme---
e.g., a space-time code with appropriate precoding--- to adapt the
transmit structures stipulated by the transmit covariances
$(\Wb_1,\ldots,\Wb_K)$. While our main interest in this paper is still
in transmit beamforming, it is worthwhile to keep such a parallel
possibility in mind, since it eliminates the need for rank-one transmit
covariances.}

%{\color{red} \it (Tsung-Hui: Can you find some references that are
%somehow related to the information theoretic model we show above? In
%fact, to researchers in interference channels, the formulation
%\eqref{eq:it} is not very surprising. I do not hope that some
%reviewers got the impression that we invent this information
%theoretic formulation.) }{\color{blue} Check thi out \cite{Fung2007TCOM}}

%%%%%%%%%%%%%%%%%%%%%%%%%%%%%%%%%%%%%%%%%%%%%%%%%%%%%%%%%%%%%%%%%%
\subsection{{Restriction Step}}
%%%%%%%%%%%%%%%%%%%%%%%%%%%%%%%%%%%%%%%%%%%%%%%%%%%%%%%%%%%%%%%%%%

Let us continue to illustrate the second step of RAR---
restriction. The relaxation step alone does not provide a convex
approximation of the main problem \eqref{eq:main}. The SDR
probabilistic constraints \eqref{eq:sdr_SINR_con} remain
intractable, although they appear to be relatively easier to handle
than the original counterparts in \eqref{eq:main_SINR_con}. The
restriction step aims to find a convex approximation of
\eqref{eq:sdr_SINR_con}, in a restrictive or conservative sense.
More precisely, in the context of the probabilistic SINR constrained problem, the restriction step entails finding a solution to the following: \\
%{To make a precise statement on this,
%we boil down the restriction step to the following abstractized problem statement:} \\

\noindent \fbox{\parbox{\linewidth}{ {\bf Challenge 1:} \ Consider the following
chance constraint:
% for a general quadratic inequality, possibly indefinite, as follows:
\begin{equation} \label{eq:ch1}
{\rm Prob}\{ \eb^H \Qb \eb + 2 {\rm Re}\{ \eb^H \rb \} + s \geq 0 \}
\geq 1 - \rho,
\end{equation}
where $\eb \in \mathbb{C}^n$ is a standard complex Gaussian vector
(i.e., $\eb \sim \mathcal{CN}( \bm{0}, \Ib_n )$), the 3-tuple $(\Qb,\rb,s)
\in \mathbb{H}^n \times \mathbb{C}^n \times \mathbb{R}$ is a set of
(deterministic) optimization variables, and $\rho \in (0,1]$ is
fixed. Find an
%convex restrictive approximation
{\it efficiently computable convex restriction} of~\eqref{eq:ch1}; i.e., find an efficiently computable convex set $\mathcal{S} \subset \mathbb{H}^n \times \mathbb{C}^n \times \mathbb{R} \times \mathbb{C}^\ell$ such that whenever $(\Qb,\rb,s,\tb) \in \mathcal{S}$, the 3-tuple $(\Qb,\rb,s) \in \mathbb{H}^n \times \mathbb{C}^n \times \mathbb{R}$ is feasible for~\eqref{eq:ch1}.
%if the constraints of the convex restriction can be represented by
%\begin{equation} \label{eq:ch1_t1}
% (\Qb,\rb,s) \in \mathcal{S}
%\end{equation}
%for some feasible set $\mathcal{S}$, then
%i) $\mathcal{S}$ must be convex and ii) the satisfiability of \eqref{eq:ch1_t1} must imply that of \eqref{eq:ch1}.
%{\color{red}\it (we may rename ``convex restrictive
%approximation''. We will think about that later. {{\rm
%[} March: we talked and think ``convex restriction'' should be fine.
%Will fix this later. {\rm ]} })}
%Specifically, if the approximation
%constraint can be represented by
%\begin{equation} \label{eq:ch1_t1}
% (\Qb,\rb,s) \in \mathcal{S}
%\end{equation}
}} ~

Note that in the construction of the convex set $\mathcal{S}$, we are allowed to include an extra optimization variable $\tb \in \mathbb{C}^\ell$, in addition to the original optimization variables $(\Qb,\rb,s) \in \mathbb{H}^n \times \mathbb{C}^n \times \mathbb{R}$.  Although the precise role of $\tb$ will depend on how the set $\mathcal{S}$ is formulated, it suffices to think of $\tb$ as a slack variable.

It is not hard to see that the SDR probabilistic constraints in~\eqref{eq:sdr_SINR_con} fall in the scope of Challenge 1.  Indeed, for
%the $i$th
each constraint in \eqref{eq:sdr_SINR_con}, the following
correspondence to \eqref{eq:ch1} can be shown:

\begin{subequations}\label{eq: definitions of qrs}
\begin{align}
\Qb = \Cb_i^{1/2} \left( \frac{1}{\gamma_i} \Wb_i - \sum_{k \neq i}
\Wb_k \right) \Cb_i^{1/2}, \quad &
\rb = \Cb_i^{1/2} \left( \frac{1}{\gamma_i} \Wb_i - \sum_{k \neq i} \Wb_k \right) \bar{\hb}_i, \\
s  = \bar{\hb}_i^H \left( \frac{1}{\gamma_i} \Wb_i - \sum_{k \neq i}
\Wb_k \right)  \bar{\hb}_i - \sigma_i^2, \quad & \rho = \rho_i.
%\notag
\end{align}
\end{subequations}

The development of convex restriction methods plays a crucial
role in RAR, and this will be our focus in
subsequent sections.
%By replacing \eqref{eq:sdr_SINR_con} with
%their respective convex restrictions,
By replacing each probabilistic constraint in~\eqref{eq:sdr_SINR_con} with
a convex restriction,
we will obtain a
convex approximation of the original probabilistic SINR
constrained problem.
Table~\ref{table_SDRs} summarizes all the RAR methods to be proposed in later sections.
Note that each RAR method is based on a different convex restriction.
%In Table \ref{table_SDRs}, we give a summary of
%all the proposed RAR methods.
Moreover, all the RAR formulations in Table~\ref{table_SDRs} are conic problems with linear matrix inequality constraints and/or
second-order cone constraints, which can be easily solved by off-the-shelf
convex optimization software~\cite{cvx}.

%The proposed RAR formulations
%%%%%%%%%%%%%%%%%%%%%%%%%%%%%%%%%%%%%%%%%%%%%%%%%%%%%%%%%%%%%%%%%%%%%%%%%%%%%%%%%%%%
\begin{table*}[]
%{\color{blue}
\begin{center}
\caption{{The proposed RAR
formulations.}}{\footnotesize
%\begin{tabular}{m{1.5cm}||m{3cm}||m{11.5cm}}
\begin{tabular}{m{.188\textwidth}||m{.76\textwidth}}
\hline
     %~~~~~~~~Method
     \centering Method
     &
     %~~~~~~~~~~~~~~~~~~~~~~~~~~~RAR formulations\\
     \hfil RAR Formulation \hfil \\
\hline\hline
    %\multirow{20}{2.5cm}{Complex Gaussian distribution}
%&
    %\begin{align*}
%    \begin{array}{ll}
%    {\bf Method~ I}: \\
%    {\bf Sphere~ bounding} \\
%    {\rm (for~ complex~ Gaussian} \\
%    {\rm CSI~ errors)}
%    \end{array}
%    \end{align*}
    \parbox{\linewidth}{
        {\bf Method I: \\ Sphere bounding} \\
        (for complex Gaussian CSI errors)
    }
%    {\bf Method I} {\bf (Sphere bound)}
&   \vspace{-0.3cm}
    \!\!\!\!\!\begin{align}\label{eq: table sphere bound}
    \begin{array}{ll}
    {\displaystyle \min_{\substack{\Wb_i \in \mathbb{H}^{N_t},
    t_i\in
    \Ral, \\ i=1,\ldots,K}}~ \sum_{i=1}^{K} {\tr}(\Wb_i)} \\
    ~~~~~~~~~~~~~{\rm s.t.} ~
    \begin{bmatrix} \!\!\!\!\!\!\!\Qb_i+t_i\Ib_{N_t} ~~~~~~\rb_i\\
    ~~~~~\rb_i^H  ~~~~~~~~s_i-t_i d^2_i
    \end{bmatrix}\succeq \bm{0},~i=1,\ldots,K,\\
    ~~~~~~~~~~~~~~~~~~~\Wb_1,\ldots,\Wb_K \succeq \zerob,
    ~t_1,\ldots,t_K\geq 0;
    \end{array}
    \end{align}
    where $\Qb_i$, $\rb_i$ and $s_i$ are defined in the same way as \eqref{eq: definitions of qrs}, and $d_i=\sqrt{\Phi_{\chi^2_{2n}}^{-1}(1-\rho_i)/2}$,
    $i=1,\ldots,K$.\vspace{0cm}\\
\hline
%\cline{2-3} &
    %\begin{align*}
%    \begin{array}{ll}
%    {\bf Method~ II:} \\
%    \text{\bf Bernstein-type } \\
%    {\bf  inequality} \\
%    {\rm (for~ complex~ Gaussian} \\
%    {\rm CSI~ errors)}
%    \end{array}
%    \end{align*}
    \parbox{\linewidth}{
        {\bf Method II: \\ Bernstein-type inequality} \\
        (for complex Gaussian CSI errors)
    }
    %{\bf Method II} {\bf (Bernstein type inequality)}
&   \vspace{-0.3cm}
    \!\!\!\begin{align}\label{eq: table bernstein inequality}
    \begin{array}{ll}
    {\displaystyle \min_{\substack{\Wb_i \in \mathbb{H}^{N_t},x_i,y_i\in
    \mathbb{R},
    \\i=1,\ldots,K}}~ \sum_{i=1}^{K} \tr(\bm{W}_i)}\\
    ~~~~~~~~~~~~~~~{\rm s.t.}~\tr(\Qb_i)-\sqrt{-2\ln(\rho_i)} \cdot x_i + \ln(\rho_i) \cdot y_i+s_i\geq 0,~i=1,\ldots,K, \\\
    ~~~~~~~~~~~~~~~~~~~~\left\|
    \begin{bmatrix}
    {\rm vec}(\Qb_i) \\
    \sqrt{2}\rb_i
    \end{bmatrix}
    \right\| \leq x_i,~i=1,\ldots,K,\\
    ~~~~~~~~~~~~~~~~~~~~y_i\Ib_{N_t}+\Qb_i\succeq \zerob,~i=1,\ldots,K, \\
    ~~~~~~~~~~~~~~~~~~~~y_1,\ldots,y_K\geq 0, ~\Wb_1,\ldots,\Wb_K
    \succeq \zerob;
    \end{array}
    \end{align}
    where $\Qb_i$, $\rb_i$ and $s_i$ are defined in the same way as \eqref{eq: definitions of
    qrs}, $i=1,\ldots,K$.   \vspace{0cm}\\
\hline
%\cline{2}
%&
%    \begin{align*}
%    \begin{array}{ll}
%    {\bf Method~ III} \\
%    {\bf(Box~ bound)}
%    \end{array}
%    \end{align*}
    \parbox{\linewidth}{
        {\bf Method III: \\ Decomposition into \\independent parts} \\
        (for complex Gaussian CSI errors)
    }
    %{\bf Method III }{\bf (Decomposition approach)}
&   \vspace{-0.3cm}
    {\!\!\!\begin{align} \label{eq: table dependence graph gaussian}
    \begin{array}{ll}
    {\displaystyle \min_{\substack{\Wb_i \in \mathbb{H}^{N_t},
    x_i,y_i\in\mathbb{R}, \\ i=1,\ldots,K}}~ \sum_{i=1}^{K} {\tr}(\Wb_i)} \\
    ~~~~~~~~~~~~~{\rm s.t.} ~s_i + \tr(\Qb_i)\geq 2\sqrt{-\ln(\rho_i)} \cdot (x_i+y_i),~i=1,\ldots,K,\\
    ~~~~~~~~~~~~~~~~~\frac{1}{\sqrt{2}}\|\rb_i\|\leq x_i,~i=1,\ldots,K,\\
    ~~~~~~~~~~~~~~~~~~v_i\left\|{\rm vec}(\Qb_i)\right\|\leq y_i,~i=1,\ldots,K,\\
    ~~~~~~~~~~~~~~~~~~\Wb_1,\ldots,\Wb_K \succeq \zerob;
    \end{array}
    \end{align}
    where $\Qb_i$, $\rb_i$ and $s_i$ are defined in the same way as \eqref{eq: definitions of
    qrs}, $\bar{\theta}_{i}$ is chosen such that $\bar{\theta}_{i}+\ln(1-\bar{\theta}_{i})=\ln(\rho_i)$, and
    $v_i= \sqrt{-\ln(\rho_i)/\bar{\theta}_{i}^2}$,
     $i=1,\ldots,K$.
    }
    \vspace{0cm}\\
\hline\hline
%    Elementwise i.i.d. bounded unknown distribution &
%    \begin{align*}
%    \begin{array}{ll}
%    {\bf Method~ IV} \\
%    {\bf(Dependence~ graph)}
%    \end{array}
%    \end{align*}
    \parbox{\linewidth}{
        {\bf Method IV: \\ Decomposition into \\independent parts} \\
        (for elementwise i.i.d. and bounded CSI errors with mean 0 and variance $\sigma_e^2$, but otherwise unknown distribution)
    }
    %{\bf Method IV }{\bf (Decomposition approach)}
&   \vspace{-0.3cm}
    {\!\!\!\begin{align} \label{eq: table dependence graph unknown}
    \begin{array}{ll}
    {\displaystyle \min_{\substack{\Wb_i \in \mathbb{H}^{N_t},
    \tb_{i}\in\mathbb{R}^{2N_t+1} \\i=1,\ldots,K}}~ \sum_{i=1}^{K} \tr(\bm{W}_i)}\\
    ~~~~~~~~~~~~~~~{\rm s.t.}~s_{i}+\sigma_e^{2} \cdot \tr(\Qb_i)
    \geq
    2\sqrt{-\ln(\rho_{i})} \cdot \sum_{\ell=0}^{2N_t}[\tb_{i}]_{\ell},~\forall i,\\
    ~~~~~~~~~~~~~~~~~~~~\sqrt{2}\|\rb_{i}\| \leq [\tb_{i}]_{0},~i=1,\ldots,K,\\
    ~~~~~~~~~~~~~~~~~~~~\left(\sum_{(j,k)\in\mathcal{A}_{\ell}}v_{jk}^{2}[\Qb_{i}]_{jk}^{2}\right)^{1/2}\leq
    [\tb_{i}]_{\ell},\\
    ~~~~~~~~~~~~~~~~~~~~~~~~~~~~~~~~~~~~\ell=1,\ldots,2N_t,~i=1,\ldots,K,\\
    ~~~~~~~~~~~~~~~~~~~~\Wb_1,\ldots,\Wb_K \succeq \zerob;
    \end{array}
    \end{align}
    where $\Qb_i$, $\rb_i$ and $s_i$ are defined in the same way as \eqref{eq: definitions of qrs2}, $\mathcal{A}_\ell$ are defined in the same way as Table \ref{ColoringTable}, and $v_{jj}=1/\sqrt{8}$ and $v_{jk}=1$ if $j\neq k$.
    }
    \vspace{0cm} \\
\hline
\end{tabular} \label{table_SDRs}}
\end{center}\vspace{-0.7cm}
%} %color{blue}
\end{table*}
%%%%%%%%%%%%%%%%%%%%%%%%%%%%%%%%%%%%%%%%%%%%%%%%%%%%%%%%%%%%%%%%%%%%%%%%%%%%%%%%%%%%

The last step of RAR is to provide a feasible beamforming
solution $(\wb_1,\ldots,\wb_K)$ to the main problem
\eqref{eq:main} by using the RAR solution $(\Wb_1,\ldots,\Wb_K)$. As is
common in all SDR-based methods, the $\Wb_i$'s obtained
from RAR may have rank higher than one. A standard way
of tackling this issue is to apply some rank-one
approximation procedure to $(\Wb_1,\ldots,\Wb_K)$ to generate a
feasible beamforming solution $(\wb_1,\ldots,\wb_K)$ to~\eqref{eq:main}; see~\cite{Luo2010_SPM}
for a review and references.
%In particular, for the problem here, the
%solution approximation can
%%in principle
%be handled by developing a
%variation of the Gaussian randomization method in
%\cite{Karipidis08}.
In our setting, we apply a Gaussian randomization procedure to a non-rank-one RAR solution.  The procedure is provided in Algorithm~\ref{alg1};
the spirit follows that of \cite{Karipidis08}, and readers are referred to \cite{Karipidis08} for an exposition of the idea.
%\footnote{ For our problem, the Gaussian
%randomization should follow the steps below. First, randomly
%generate beamformer directions $\bm{u}_1,\ldots,\bm{u}_K$, $\|
%\bm{u}_i \|^2 = 1$ for all $i$, based on the RAR solution
%$(\Wb_1,\ldots,\Wb_K)$. Second, fix the beamformer directions and
%solve a power control problem; i.e., substitute $\bm{W}_i = p_i
%\bm{u}_i \bm{u}_i^H$ into the RAR problem, where $p_i$ denotes the
%power, and optimize it with respect to $p_1,\ldots,p_K \geq 0$.
%Third, if the power control problem is feasible, then combine the
%obtained powers and the beamformer directions to form a feasible
%approximate solution of the main problem. Last, repeat the above
%steps multiple times, and pick the one that yields the lowest power.
%%{\color{red} \it (I am not sure if this is too long, but at least the idea is there...)}
We should point out that obtaining a feasible RAR
solution does not imply that we can always generate a feasible
solution to the main problem~\eqref{eq:main}.  This issue has also been
identified before in the context of multigroup multicast beamforming with
perfect CSI~\cite{Karipidis08}. However, if the RAR solution happens
to give rank-one $\Wb_i$ for all $i$, then we can simply solve the
rank-one decomposition $\Wb_i = \wb_i \wb_i^H$ and output the
corresponding $(\wb_1,\ldots,\wb_K)$ as the approximate beamforming
solution.  For such instances, it can be easily verified that
$(\wb_1,\ldots,\wb_K)$ is already feasible for the main problem
\eqref{eq:main}.
%Hence, those ``good'' instances have simple
%approximate solution generations,
%and assure feasibility of the approximate solution generated
%will be feasible as far as the RAR problem is feasible.
Rather surprisingly, we found that the proposed RAR methods returned rank-one
solutions in almost all the simulation trials we ran\footnote{A similar phenomenon was observed in a different problem setting, namely that of the worst-case SINR constrained design~\cite{ZhengWongNg_2008}.}.
%{\color{blue} usually}
%found in simulations--- the proposed
%RAR methods returned rank-one solutions in
%{\color{blue} almost all of}
%the simulation trials
%we ran.
%As it turns out, this mysterious yet appealing
%result has also been noticed in a worst-case robust unicast
%SINR constrained formulation~\cite{ZhengWongNg_2008}.
Such empirical finding provides another interesting implication when we
consider the information-theoretic interpretation in the last
subsection: Since the RAR methods are essentially the same as
convex restrictions of the outage-based rate
optimization problem in \eqref{eq:it}, the numerical observation that
RAR solutions are almost always rank-one somehow hints that
transmit beamforming may inherently be an optimal physical-layer
scheme, at least for the outage-based unicast multiuser MISO
downlink scenario considered here.

%\begin{center}
%\begin{minipage}{.75\textwidth}%

\begin{algorithm}[htbp]
%{\color{blue}
\caption{Gaussian randomization procedure for RAR} \label{alg1}
%\begin{algorithmic}[H]
\SetAlgoNoLine
\SetKwInOut{Given}{Given} \SetKwInOut{Output}{Output}

\Given{A number of randomizations $L$, and
 an optimal solution $( \Wb_1^\star,\dots,\Wb_K^\star)$ to an employed RAR formulation.}

\For{$\ell=1,\ldots,L$}{
    generate random vectors $\wb_i^{(\ell)} \sim \mathcal{CN}( \bm{0}, \Wb_i^\star)$,
    $i=1,\ldots,K$\;
    set beam directions $\ub_i^{(\ell)}={\wb_i^{(\ell)}}/{\|\wb_i^{(\ell)}\|}$, $i=1,\ldots,K$\;
    let $p_1^{(\ell)},\ldots,p_K^{(\ell)}$ be beam powers and obtain $p_1^{(\ell)},\ldots,p_K^{(\ell)}$ as follows:
    substitute $\bm{W}_i = p_i \ub_i^{(\ell)} (\ub_i^{(\ell)})^H$, $i=1,\ldots,K$, into the RAR problem,
    solve the problem with respect to $p_1,\ldots,p_K \geq 0$,
    and set $p_1^{(\ell)},\ldots,p_K^{(\ell)}$ as its solution if the problem is feasible; also, set $P^{(\ell)}$ to be the associated optimal objective value if the problem is feasible; otherwise set $P^{(\ell)} = \infty$ \;
}

$\ell^\star=\arg\min_{\ell=1,\ldots,L} P^{(\ell)}$.

\Output{$\hat\wb_i^\star=\sqrt{p_i^{(\ell^\star)}}\ub_i^{(\ell^\star)}$, $i=1,\ldots,K$, as an approximate solution to the main problem \eqref{eq:main}.}

%}
\end{algorithm}

%\end{minipage}%
%\end{center}

%%%%%%%%%%%%%%%%%%%%%%%%%%%%%%%%%%%%%%%%%%%%%%%%%%%%%%%%%%%%%%%%%%%%%
%\begin{table}[h]
%\caption{Gaussian randomization procedure for the proposed RAR formulations in Table \ref{table_SDRs}}
%\begin{center} \small
%%\vspace{\baselineskip}
%\parbox{\linewidth}{\footnotesize
%\rule{\linewidth}{1pt}
%
%\begin{Ventry}{ {\bf Step 4.} }
%
%\item[{\bf Given}] a number of randomizations $L$, and an optimal solution $( \Wb_1^\star,\dots,\Wb_K^\star)$ of an employed RAR formulation.
%
%
%\item[{\bf Step 1.}] For each $i=1,\dots,K$, generate a set of $L$ random vectors
%$\wb_i^{(\ell)},$ $\ell=1,\ldots,L$, from the complex Gaussian
%distribution $\mathcal{CN}({\bf 0}, \Wb_i^\star)$.
%
%\item[{\bf Step 2.}] For $\ell=1,\ldots,L$, let
%$\ub_i^{(\ell)}={\wb_i^{(\ell)}}/{\|\wb_i^{(\ell)}\|}$ for all
%$i=1,\dots,K,$ and solve a power control problem by substituting each $\bm{W}_i = p_i
%\ub_i^{(\ell)} (\ub_i^{(\ell)})^H$ into the RAR problem, where $p_1,\ldots,p_K \geq 0$ denote the
%powers to be optimized.
%For each $\ell$, let $(p_1^{(\ell)},\ldots,p_K^{(\ell)})$ be the obtained optimal powers and
%$P^{(\ell)}$ be the associated optimal objective value.
%
%\item[{\bf Step 3.}]  Let $$\ell^\star=\arg~\min_{\ell=1,\ldots,L} P^{(\ell)},$$
%and output $
%\hat\wb_i^\star=\sqrt{p_i^{(\ell^\star)}}\ub_i^{(\ell^\star)},~i=1,\ldots,K,~
%$ as an approximate solution to problem \eqref{eq:main}.
%\end{Ventry}
%\rule{\linewidth}{1pt} }
%\end{center}
%
%\vspace*{-\baselineskip}\label{Table specialized IPA}\vspace{-0.3cm}
%\end{table}
%%%%%%%%%%%%%%%%%%%%%%%%%%%%%%%%%%%%%%%%%%%%%%%%%%%%%%%%%%%%%%%%%%%%%

As a summary to the solution approximation aspect discussed above,
in most cases a simple rank-one decomposition $\Wb_i =
\wb_i \wb_i^H$ of the RAR solution suffices to produce a feasible solution
$(\wb_1,\ldots,\wb_K)$ to the main problem~\eqref{eq:main}.
The more complicated solution approximation procedure in Algorithm~\ref{alg1}, proposed for instances where the RAR solution is not of rank one, is rarely needed in our empirical experience.

%%%%%%%%%%%%%%%%%%%%%%%%%%%%%%%%%%%%%%%%%%%%%%%%%%%%%%%%%%%%%%%%%%
\section{RAR Method I: Sphere Bounding}\label{sec: sphere bounding}
%%%%%%%%%%%%%%%%%%%%%%%%%%%%%%%%%%%%%%%%%%%%%%%%%%%%%%%%%%%%%%%%%%

In this section we describe our first convex restriction
method for Challenge 1.
%, thereby establishing an RAR
%method for the probabilistic SINR constrained problem.
%% under the complex Gaussian CSI error model.
%The first convex restriction
The method is based on two
key ingredients. The first is the following lemma:
\begin{Lemma} \label{lem:worst}
Consider Challenge 1. Suppose that we have a set $\mathcal{B}
\subset \mathbb{C}^n$ that satisfies
\begin{equation} \label{eq:lemma1_t1}
{\rm Prob}\{ \bm{e} \in \mathcal{B} \} \geq 1 - \rho.
\end{equation}
Then, the following implication holds:
\begin{equation} \label{eq:lemma1_t2}
\begin{array}{c}
\bm{\delta}^H \bm{Q} \bm{\delta} + 2 {\rm Re}\{ \bm{\delta}^H \bm{r} \} + s \geq 0, \\
\text{for all~} \bm{\delta} \in \mathcal{B}
\end{array} \Longrightarrow
\text{Eq.~(8) in Challenge 1 holds.}
\end{equation}
\end{Lemma}
The proof of Lemma~\ref{lem:worst} is simple and is given as
follows. Let $p(\bm{e})$ denote the probability density function of
$\bm{e}$. Suppose that \eqref{eq:lemma1_t1} and the left-hand side
(LHS) of \eqref{eq:lemma1_t2} hold.  Then, we have the following chain:
\begin{align*}
{\rm Prob}\{ \bm{e}^H \bm{Q} \bm{e} + 2 {\rm Re}\{ \bm{r}^H \bm{e}
\} + s \geq 0 \}
& = \int_{ \bm{e}^H \bm{Q} \bm{e} + 2 {\rm Re}\{ \bm{r}^H \bm{e} \} + s \geq 0 } p(\bm{e}) d\bm{e} \\
& \geq \int_{ \bm{e} \in \mathcal{B} } p(\bm{e}) d\bm{e} \\
& \geq 1 - \rho.
\end{align*}
Hence, Eq.~\eqref{eq:ch1} is satisfied.

Lemma 1 suggests that we can approximate the chance constraint in
\eqref{eq:ch1} in a conservative (or restrictive) fashion by using the
worst-case deterministic constraint on the LHS of
\eqref{eq:lemma1_t2}. Moreover, it can be easily seen that the same
idea applies to general chance constraints; i.e.,
%the inequality function in the probability does not necessarily need to be quadratic.
the quadratic functions in \eqref{eq:ch1} and \eqref{eq:lemma1_t2}
may be replaced by any arbitrary function.  Such an insight (i.e., using
worst-case deterministic constraints to approximate (general) chance
constraints) have been alluded to or used in many different contexts; e.g.,
\cite{Ben-tal2000,Bertsimas06} in optimization. Here, we are
interested in the chance constraint in
\eqref{eq:ch1}, which involves a quadratic function of the standard complex
Gaussian vector $\bm{e}$.
In our method, we choose $\mathcal{B}$ to be a spherical set; i.e.,
\begin{equation}
\mathcal{B}= \{ \bm{\delta} \in \mathbb{C}^n \mid \|
\bm{\delta} \| \leq d \}, \label{ball uncertainty}
\end{equation}
where $d$ is the sphere radius. It can be shown that by choosing
%\begin{align}
$$ d = \sqrt{ \frac{ \Phi_{\chi^2_{2n}}^{-1} (1-\rho) }{2} }, $$
%\label{radius}
%\end{align}
where $\Phi_{\chi^2_m}^{-1} ( \cdot )$ is the inverse
cumulative distribution function of the (central)
Chi-square random variable with $m$ degrees of freedom,
Eq.~\eqref{eq:lemma1_t1} is satisfied.

The second ingredient is the so-called $\mathcal{S}$-lemma, which enables us
to turn the infinitely many constraints on the LHS of \eqref{eq:lemma1_t2} into a set of tractable constraints.  The $\mathcal{S}$-lemma is given as follows:
\begin{Lemma}[ {$\mathcal{S}$-lemma \cite{BK:Ben-Tal01}} ]
Let $f_i(\bm{x})= \bm{x}^H \bm{Q}_i \bm{x} + 2 {\rm Re}\{ \bm{x}^H
\bm{r}_i \} + s_i$ for $i=0,1$, where $\bm{x} \in \mathbb{C}^n$ and
$(\bm{Q}_i, \bm{r}_i, s_i) \in \mathbb{H}^n \times \mathbb{C}^n
\times \mathbb{R}$ for $i=0,1$. Suppose that there exists an
$\hat{\bm{x}} \in \mathbb{C}^n$ satisfying $f_1(\hat{\bm{x}}) < 0$. Then, the
following statements are equivalent:
\begin{itemize}
\item[1.] $f_0(\bm{x}) \geq 0$ for all $\bm{x} \in \mathbb{C}^n$
satisfying $f_1(\bm{x}) \leq 0$.
\item[2.] There exists a $t \geq 0$ such that
\begin{equation} \label{eq: S_lemma_2}
\begin{bmatrix}
\bm{Q}_0 & \bm{r}_0 \\ \bm{r}_0^H & s_0
\end{bmatrix} + t
\begin{bmatrix}
\bm{Q}_1 & \bm{r}_1 \\ \bm{r}_1^H & s_1
\end{bmatrix} \succeq \bm{0}.
\end{equation}
\end{itemize}
\end{Lemma}
%Applying the $\mathcal{S}$-lemma to the LHS of
%\eqref{eq:lemma1_t2} with $\mathcal{B}$ defined as in \eqref{ball
%uncertainty}, we yield the following convex constraints:
%\begin{equation}
%\begin{bmatrix}
%\bm{Q}+ \lambda \bm{I} & \bm{r} \\ \bm{r}^H & s- \lambda d^2
%\end{bmatrix}  \succeq \bm{0}, \qquad \lambda \geq 0
%\end{equation}
%We therefore have built a convex restriction method
%for Challenge 1. The method is summarized as follows:
By the $\mathcal{S}$-lemma, the LHS of \eqref{eq:lemma1_t2}, with $\mathcal{B}$ given by \eqref{ball uncertainty}, can be equivalently represented by an LMI of the form \eqref{eq: S_lemma_2}, where $(\bm{Q}_0,\bm{r}_0,s_0) = ( \bm{Q}, \bm{r}, s )$ and
$(\bm{Q}_1,\bm{r}_1,s_1) = ( \Ib_n, \bm{0}, -d^2 )$.
We therefore have built a convex restriction for Challenge 1.
To summarize, we have the following:

~ \\
\noindent \fbox{\parbox{\linewidth}{ {\bf Method I for
Challenge 1 (Sphere bounding):} \ The following feasibility
problem is a convex restriction of \eqref{eq:ch1} in
Challenge 1:
\begin{align*}
\text{Find}&~~ \Qb,\rb,s,t \\
\text{s.t.}&~~
\begin{bmatrix}
\bm{Q}+ t \Ib_n & \bm{r} \\ \bm{r}^H & s- t d^2
\end{bmatrix}  \succeq \bm{0},\notag\\
&~~t \geq 0,%\notag
\end{align*}
where $d = \sqrt{\Phi_{\chi^2_{2n}}^{-1} (1-\rho) /2 }$.
}} ~

By first applying SDR and then Method I to the probabilistic SINR
constrained problem~\eqref{eq:main}, we obtain the RAR
formulation \eqref{eq: table sphere bound} in Table \ref{table_SDRs}.
%, where it
%is defined that
%\begin{subequations}\label{eq: definitions of qrs}
%\begin{align}
% \Qb_i&=\Cb_i^{1/2}\left(\frac{1}{\gamma_i}\Wb_i-\sum_{k\neq
%    i}\Wb_k\right)\Cb_i^{1/2}, \\
% \rb_i&=\Cb_i^{1/2}\left(\frac{1}{\gamma_i}\Wb_i-\sum_{k\neq
%    i}\Wb_k\right)\bar{\hb}_i, \\
% s_i&=\bar{\hb}_i^H\left(\frac{1}{\gamma_i}\Wb_i-\sum_{k\neq
%    i}\Wb_k\right)\bar{\hb}_i-\sigma_i^2,
%\end{align}
%\end{subequations}for $i=1,\ldots,K$.
Interestingly, this formulation turns out to be similar to that of
the worst-case robust SDR problem considered in \cite{ZhengWongNg_2008}.
However, it should be noted that the prior work does not consider
outage probability constraints.  Moreover, we show a way of using the
worst-case robust formulation to deal with the probabilistic SINR constrained
problem.  Finally, by incorporating the bisection scheme proposed
in \cite{Shenouda2008}, which will be considered
in our simulations in Section \ref{sec: simulation results}
(Example 3), we will be able to further improve the performance of the
sphere bounding RAR method.

%%%%%%%%%%%%%%%%%%%%%%%%%%%%%%%%%%%%%%%%%%%%%%%%%%%%%%%%%%%%%%%%%%
\section{Probability Inequality Approaches}\label{sec: probability inequality approaches}
%%%%%%%%%%%%%%%%%%%%%%%%%%%%%%%%%%%%%%%%%%%%%%%%%%%%%%%%%%%%%%%%%%

The reader may notice that the development of Method I is strongly
motivated by the worst-case robust optimization paradigm. Indeed,
the problem on the LHS of the implication \eqref{eq:lemma1_t2} is
precisely a robust feasibility problem with uncertainty set
$\mathcal{B}$. By choosing $\mathcal{B}$ judiciously, it is shown
that the violation probability ${\rm Prob}\{ \bm{e}^H \bm{Q} \bm{e} + 2 {\rm Re}\{ \bm{r}^H \bm{e} \} + s < 0 \}$ can be controlled, and the resulting
robust feasibility problem is a convex restriction of
\eqref{eq:ch1}. However, this approach has an intrinsic
drawback, namely, it is difficult to define and analyze an
uncertainty set $\mathcal{B}$ other than those that have very simple
geometry, such as the spherical set considered in the previous
section. Consequently, it is not clear whether there exist other
choices of $\mathcal{B}$ that would lead to better convex
restrictive approximations.

As it turns out, one can circumvent the above drawback by using
analytic upper bounds on the violation probability to construct
efficiently computable convex restrictions of \eqref{eq:ch1}.
Specifically, suppose that we have an efficiently computable
convex function $f(\bm{Q},\bm{r},s,\bm{t})$, where
$\bm{t}$ is an extra optimization variable, such that
\begin{align} \label{eq: probability1}
{\rm Prob}\{ \eb^H \Qb \eb + 2 {\rm Re}\{ \eb^H \rb \} + s <  0 \} \le f(\bm{Q},\bm{r},s,\bm{t}).
\end{align}
Then, the constraint
\begin{align} \label{eq: restriction condition}
f(\bm{Q},\bm{r},s,\bm{t}) \le \rho
\end{align}
is, by construction, a convex restriction of \eqref{eq:ch1}.
An upshot of this approach is that
there are many available techniques for constructing such upper
bounds, and each of those bounds yields a convex restriction
of \eqref{eq:ch1}. Moreover, it is known \cite[Chapter
4]{BEN09} that under some fairly mild conditions,
every convex restriction corresponds to
a robust feasibility problem with a suitably defined uncertainty
set. Thus, the above approach can be viewed as an enhancement of Method
I, in the sense that it provides a handle on more sophisticated
uncertainty sets that are difficult to construct directly.

%%%%%%%%%%%%%%%%%%%%%%%%%%%%%%%%%%%%%%%%%%%%%%%%%%%%%%%%%%%%%%%%%%
\subsection{Method II: Bernstein-Type Inequality}
%%%%%%%%%%%%%%%%%%%%%%%%%%%%%%%%%%%%%%%%%%%%%%%%%%%%%%%%%%%%%%%%%%

Let us now illustrate the above approach by showing how a
Bernstein-type inequality for Gaussian quadratic forms can be used
to construct a convex restriction of \eqref{eq:ch1}.
{Our approach relies on the following lemma due to Bechar
\cite{Bechar2009}:

\begin{Lemma}\label{lemma Bernstein inequality}
Let $\eb\sim \mathcal{CN}(\zerob,\Ib_n)$, $\Qb\in\mathbb{H}^{n}$ and
$\rb\in\mathbb{C}^{n}$. Then, for any $\eta >0$, we have
\begin{align}\label{eq: bernstein inequality}
\Prob\{\eb^H \Qb \eb + 2 {\rm Re}\{ \eb^H \rb \} \geq T(\eta)
\} \geq 1- e^{-\eta},
\end{align}
where the function $T: \mathbb{R}_{++} \rightarrow \mathbb{R} $ is
defined by
\begin{align}
T(\eta)=\tr(\Qb) - \sqrt{2\eta}\sqrt{\|\Qb\|_{F}^2 + 2\|\rb\|^2}
-\eta \lambda^{+}(\Qb),
\end{align}
with $\lambda^{+}(\Qb)=\max\{\lambda_{\max}(-\Qb),0\}$.
\end{Lemma}
Lemma \ref{lemma Bernstein inequality} is obtained by extending the
corresponding result in \cite{Bechar2009} for quadratic
forms of real-valued Gaussian random variables. The inequality in
\eqref{eq: bernstein inequality} is a so-called Bernstein-type
inequality\footnote{Roughly speaking, a Bernstein-type inequality is
one which bounds the probability that a sum of random variables
deviates from its mean.  The famous Markov inequality, Chebyshev inequality
and Chernoff bounds can all be viewed as instances of Bernstein-type
inequalities.},
which bounds the probability that the quadratic form $\eb^H \Qb \eb
+ 2 {\rm Re}\{ \eb^H \rb \}$ of complex Gaussian random variables
deviates from its mean $\tr(\Qb)$.

Since $T(\eta)$ is monotonically decreasing, its inverse mapping $T^{-1}:
\mathbb{R} \rightarrow \mathbb{R}_{++}$ is well defined.  In particular,
the Bernstein-type inequality in \eqref{eq: bernstein
inequality} can be expressed as
\begin{align}\label{eq: bernstein inequality2}
\Prob\{\eb^H \Qb \eb + 2 {\rm Re}\{ \eb^H \rb \} + s \geq
0\} \geq 1- e^{-T^{-1}(-s)}.
\end{align}
As discussed in \eqref{eq: probability1} and \eqref{eq: restriction
condition}, the constraint $e^{-T^{-1}(-s)} \leq \rho$, or
equivalently,
\begin{align}
\tr(\Qb)-\sqrt{-2\ln(\rho)}\sqrt{\|\Qb\|_{F}^2+2\|\rb\|^2} + \ln(\rho) \cdot
\lambda^{+}(\Qb)+ s \geq 0 \label{conservative approach 1}
\end{align}
serves as a sufficient condition for achieving \eqref{eq:ch1}.

While it is not obvious at this stage whether \eqref{conservative
approach 1} is convex in $(\Qb,\rb,s)$ or not, a crucial observation
is that \eqref{conservative approach 1} can be equivalently
represented by the following system of convex conic inequalities:
\begin{subequations}\label{reformulated constraints 1}
\begin{align}
&~\tr\left(\Qb\right)-\sqrt{-2\ln(\rho)} \cdot t_1 + \ln(\rho) \cdot
t_2 + s \geq 0, \label{constraint_BT_11}\\
&~\sqrt{\|\Qb\|_{F}^2+2\|\rb\|^2} \leq t_1, \label{constraint_BT_12}\\
&~t_2 \Ib_n+\Qb\succeq \zerob, \label{constraint_BT_13}\\
&~t_2 \geq 0, \label{constraint_BT_14}
\end{align}
\end{subequations}
where $t_1,t_2\in \mathbb{R}$ are slack variables. Therefore,
formulation~\eqref{reformulated constraints 1} is an
efficiently computable convex restriction of \eqref{eq:ch1}.
We now summarize the Bernstein-type inequality method as follows:

~ \\
\noindent \fbox{\parbox{\linewidth}{ {\bf Method II for Challenge 1
(Bernstein-type inequality method):} \ The following feasibility
problem is a convex restriction of \eqref{eq:ch1} in
Challenge 1:
\begin{align*}
 \text{Find}&~~ \Qb,\rb,s,\tb \\
 \text{s.t.}&~~
    \tr\left(\Qb\right)-\sqrt{-2\ln(\rho)}\cdot t_1 + \ln(\rho) \cdot
    t_2 +s \geq 0, \notag\\
    &~~\sqrt{\|\Qb\|_{F}^2+2\|\rb\|^2} \leq t_1,\notag\\
    &~~t_2\Ib_n+\Qb\succeq \zerob, \notag\\
    &~~t_2\geq 0. \notag
\end{align*}
}} ~ %\\

Upon applying Method II to \eqref{eq:sdr}, we obtain the RAR
formulation \eqref{eq: table bernstein inequality} in
Table \ref{table_SDRs}. %for problem \eqref{eq:main}.
As can be easily seen from the formulations \eqref{eq: table sphere bound}
and \eqref{eq: table bernstein inequality}, the latter has a
more complex constraint set and thus a higher computational
complexity in general. However, it will be shown later that
the Bernstein-type inequality method~\eqref{eq: table bernstein inequality}
exhibits better
approximation performance than the sphere bounding method.}

%%%%%%%%%%%%%%%%%%%%%%%%%%%%%%%%%%%%%%%%%%%%%%%%%%%%%%%%%%%%%%%%%%
\subsection{Method III: Decomposition into Independent Parts}
%%%%%%%%%%%%%%%%%%%%%%%%%%%%%%%%%%%%%%%%%%%%%%%%%%%%%%%%%%%%%%%%%%

For both the sphere bounding and Bernstein-type inequality
methods, the resulting convex restrictions of
\eqref{eq:ch1} contain linear matrix inequality constraints.  As
such, they could be computationally costly when the problem size is
large.  It turns out that one can also develop a convex restriction
of \eqref{eq:ch1} that contains only second-order
cone constraints.  The resulting formulation can thus be solved
more efficiently than those developed using the sphere bounding
or Bernstein-type inequality method.  The idea is to first
decompose the sum $\eb^H \Qb \eb + 2 {\rm Re}\{ \eb^H \rb \} + s$
into several parts, each of which is a sum of independent random
variables.  Then, one bounds the moment generating function of each
of those parts and stitch the results together to obtain an analytic
upper bound on the violation probability~\cite{CSW11}.  To
illustrate this approach, let $\Qb = \Ub \Lambdab \Ub^H$ be the
spectral decomposition of $\Qb$, where $\Lambdab =
\Diag(\lambda_1,\ldots,\lambda_n)$ and $\lambda_1,\ldots,\lambda_n$
are the eigenvalues of $\Qb$.  Since $\eb\sim
\mathcal{CN}(\zerob,\Ib_n)$ and $\Ub^H$ is unitary, we have
$\tilde{\eb} = \Ub^H\eb \sim \mathcal{CN}(\zerob,\Ib_n)$.  Thus, we
can write
$$ \Psi = \eb^H \Qb \eb + 2 {\rm Re}\{ \eb^H \rb \} = \tilde{\eb}^H \Lambdab \tilde{\eb} + 2 {\rm Re}\{ \eb^H \rb \} = \Psi_q + \Psi_l. $$
Now, observe that both
$$ \Psi_q = \tilde{\eb}^H \Lambdab \tilde{\eb} = \sum_{\ell=1}^n \lambda_{\ell} |e_{\ell}|^2 \quad\mbox{and}\quad \Psi_l = 2{\rm Re}\{ \eb^H \rb \} =  2\sum_{\ell=1}^n \left( {\rm Re}\{r_{\ell}\}{\rm Re}\{e_{\ell}\} + {\rm Im}\{r_{\ell}\}{\rm Im}\{e_{\ell}\} \right) $$
are sums of independent random variables.  Moreover, it can be shown
that for any fixed $\bar{\theta} < 1$,
\begin{eqnarray*}
& & {\mathbb E}\left\{ \exp\left( \theta(|e_{\ell}|^2-1) \right) \right\} = \frac{\exp(-\theta)}{1-\theta} \le \exp\left( v^2 \theta^2 \right), \\ %\quad\mbox{for } \theta \le \bar{\theta} = 0.999, \\
\noalign{\medskip} & & {\mathbb E}\left\{ \exp\left( \theta \cdot
2{\rm Re}\{ e_{\ell} \} \right) \right\} = {\mathbb E}\left\{
\exp\left( \theta \cdot 2{\rm Im}\{ e_{\ell} \} \right) \right\} =
\exp\left(\frac{1}{2} \theta^2 \right) \quad\mbox{for } \theta \in
{\mathbb R},
\end{eqnarray*}
where $v=\left(
-\left(\bar{\theta}+\ln\left(1-\bar{\theta}\right)\right)/
\bar{\theta}^2 \right)^{1/2} < \infty$.  Thus, for any $p_1,p_2>0$
such that $p_1+p_2=1$, the chain of inequalities
\begin{align}
   {\mathbb E}\left\{ \exp(u(\tr(\Lambdab) - \Psi)) \right\} =&~ {\mathbb E}\left\{ \exp\left( p_1\cdot\frac{(-u)}{p_1}(\Psi_q - \tr(\Lambdab)) + p_2\cdot\frac{(-u)}{p_2}\Psi_l \right) \right\} \nonumber \\
   %\noalign{\medskip}
   \leq &~ p_1 {\mathbb E}\left\{ \exp\left( -\frac{u}{p_1}(\Psi_q-\tr(\Lambdab)) \right)\right\} + p_2 {\mathbb E}\left\{ \exp\left(-\frac{u}{p_2} \Psi_l \right)\right\} \label{eq:jensen} \\
   %\noalign{\medskip}
   =&~ p_1 \prod_{\ell=1}^n {\mathbb E}\left\{ \exp\left( -\frac{u}{p_1} \lambda_{\ell}(|e_{\ell}|^2-1) \right) \right\} \label{eq:indep} \\
   %\noalign{\medskip}
   &+ p_2 \prod_{\ell=1}^n {\mathbb E}\left\{ \exp\left( -\frac{u}{p_2} 2{\rm Re}\{r_{\ell}\}{\rm Re}\{e_{\ell}\} \right) \right\} {\mathbb E}\left\{ \exp\left( -\frac{u}{p_2} 2{\rm Im}\{r_{\ell}\}{\rm Im}\{e_{\ell}\} \right) \right\} \nonumber \\
   %\noalign{\medskip}
   \leq &~ p_1 \exp\left( \sum_{\ell=1}^n v^2 \frac{u^2\lambda_{\ell}^2}{p_1^2} \right) + p_2 \exp\left( \sum_{\ell=1}^n \frac{1}{2} \left( \frac{u^2{\rm Re}\{r_{\ell}\}^2}{p_2^2} + \frac{u^2{\rm Im}\{r_{\ell}\}^2}{p_2^2} \right) \right) \label{eq:mgf-bd}
\end{align}
holds whenever $-u\lambda_{\ell}/p_1 < \bar{\theta}$ for
$\ell=1,\ldots,n$, where \eqref{eq:jensen} follows from Jensen's
inequality and \eqref{eq:indep} follows from the independence of the
random variables in $\Psi_q$ and $\Psi_l$.  By setting
$$ c_1 = v^2 \sum_{\ell=1}^n \lambda_{\ell}^2, \quad c_2 = \frac{1}{2} \|\rb\|^2, \quad T = \sqrt{c_1}+\sqrt{c_2}, \quad p_1 = \frac{\sqrt{c_1}}{T}, \quad p_2 = \frac{\sqrt{c_2}}{T}, $$
we see from~\eqref{eq:mgf-bd} that the inequality
$$ {\mathbb E}\left\{ \exp(u(\tr(\Lambdab) - \Psi)) \right\} \le p_1 \cdot \exp\left( u^2T^2 \right) + p_2 \cdot \exp\left( u^2T^2 \right) = \exp\left( u^2T^2 \right) $$
holds whenever $|u|T < \bar{\theta}v$.  In particular, by Markov's
inequality, it can be shown that for any $\zeta>0$,
\begin{eqnarray}
   \Prob\{ \tr(\Lambdab) - \Psi \geq \zeta \} &\le& \inf_{0< u < \bar{\theta}v/T} \Big\{ \exp(-u\zeta) \cdot {\mathbb E}\left\{ \exp(u(\tr(\Lambdab) - \Psi)) \right\} \Big\} \nonumber \\
   \noalign{\medskip}
   &=& \left\{ \begin{array}{l@{\qquad}l}
   \displaystyle{\exp\left( -\frac{\zeta^2}{4T^2} \right)} & \mbox{for } 0 < \zeta < 2\bar{\theta}vT , \\
   \noalign{\medskip}
   \displaystyle{\exp\left( -\frac{\bar{\theta}v\zeta}{T} + (\bar{\theta}v)^2 \right)} & \mbox{for } \zeta \ge 2\bar{\theta}vT.
   \end{array} \label{eq:tail-prob}
\right.
\end{eqnarray}
Now, set $\zeta = s + \tr(\Lambdab)$.  Then, the LHS of
\eqref{eq:tail-prob} becomes
$$ \Prob\{\Psi + s \le 0\} = \Prob\left\{ \eb^H \Qb \eb + 2 {\rm Re}\{ \eb^H \rb \} + s \le 0 \right\}. $$
In particular, by imposing the constraint that the right-hand side
of \eqref{eq:tail-prob} is less than $\rho$ and using the fact that
$\tr(\Lambdab) = \tr(\Qb)$ and $\sum_{\ell=1}^n \lambda_{\ell}^2 =
\|\Qb\|_F^2$, followed by some tedious derivations (see~\cite{CSW11} for details), we obtain the following method for Challenge 1.

~ \\
\noindent \fbox{\parbox{\linewidth}{ {\bf Method III for Challenge 1
(Decomposition into Independent Parts):} Given a parameter
$\bar{\theta}<1$, let
$$ v=\left( -\frac{\bar{\theta}+\ln\left(1-\bar{\theta}\right)}{\bar{\theta}^2} \right)^{1/2} $$
and
   \[
    \mu = \left\{
    \begin{array}{ll}
    2\sqrt{-\ln(\rho)}, & \mbox{if~$\bar{\theta}v>\sqrt{-\ln(\rho)}$},\\
    \bar{\theta}v-\frac{\ln(\rho)}{\bar{\theta}v}, & \mbox{otherwise}.
    \end{array}
    \right.
    \]
Then, the following feasibility problem is a convex restriction
of \eqref{eq:ch1} in Challenge 1:
\begin{align} \label{eq: method 3}
 \text{Find}&~~ \Qb,\rb,s,\tb \\
 \text{s.t.}&~~
    s+\tr\left(\Qb\right)-\mu (t_1+t_2) \geq 0, \notag \\
    &~~\frac{1}{\sqrt{2}} \|\rb\| \le t_1,\notag\\
    &~~ v \|\Qb\|_F \le t_2.  \notag
\end{align}
}} ~ %\\

%\vspace{-1cm}

Observe that in Method III, we have the flexibility to choose the
parameter $\bar{\theta}$.  Ideally, $\bar{\theta}$ should be chosen
so that both $\mu$ and $v$ are small, since then the constraints in
\eqref{eq: method 3} are easier to satisfy.  However, as can be seen
from the definition, $\mu$ and $v$ cannot be chosen independently of
each other.  Our simulation results suggest that it is better to
have a smaller value of $\mu$; i.e., choose $\bar{\theta}$ (and
hence $v$) such that $\mu = 2\sqrt{-\ln(\rho)}$.  Specifically, for
a given $\rho\in(0,1)$, we choose $\bar{\theta}$ such that $v$ is
minimized and $\mu = 2\sqrt{-\ln(\rho)}$.  This can be achieved by
solving $\bar{\theta} v = \sqrt{-\ln(\rho)}$, or equivalently,
\begin{equation} \label{eq:opt-theta}
   \bar{\theta} + \ln(1-\bar{\theta}) = \ln(\rho),
\end{equation}
which can be done numerically.  We remark that for small values of
$\rho$ (say, $\rho \in (0,0.2)$), the solution $\bar{\theta}$ to
\eqref{eq:opt-theta} can be approximated by
$$ \bar{\theta} \approx 1 - \exp(\ln(\rho)-1). $$

%%%%%%%%%%%%%%%%%%%%%%%%%%%%%%%%%%%%%%%%%%%%%%%%%%%%%%%%%%%%%%%%%%
\subsection{{Variation on a Theme: i.i.d. Bounded CSI Errors with Unknown Distribution via the Decomposition
Approach}}
%%%%%%%%%%%%%%%%%%%%%%%%%%%%%%%%%%%%%%%%%%%%%%%%%%%%%%%%%%%%%%%%%%

An advantage of the decomposition approach outlined above is that it
can be applied to cases where the distribution of the random vector
$\eb$ is not Gaussian.  As an illustration, let us generalize the
setting considered in the previous section and develop an RAR method
for handling the {\it elementwise i.i.d.~bounded support model with unknown distribution}.
In this model,
the real and imaginary parts of the CSI error vector $\eb_i$
are assumed to be independent and have i.i.d. components.  Each
component has zero mean and is supported on, say,
$[-\epsilon_i,\epsilon_i]$, where $\epsilon_i>0$. Again, we pose the
restriction step in RAR as the following generic challenge:

\begin{center}
\fbox{\parbox[]{0.97 \linewidth}{ {\bf Challenge 2:} Consider the
following chance constraint:
\begin{align}\label{eq:ch3}
  \Prob\left\{\eb^T\Qb\eb + 2\eb^T\rb + s \geq 0\right\} \geq 1-\rho,
\end{align}
where $\eb\in \mathbb{R}^{n}$ is a mean-zero random vector
supported on $[-\sqrt{3},\sqrt{3}]^n$ with independent components, the 3-tuple $(\Qb,\rb,s)\in
\mathbb{S}^n \times \mathbb{R}^{n} \times \mathbb{R}$ is a set of
optimization variables, and $\rho\in (0,1]$ is fixed. Find an efficiently
computable convex restriction of \eqref{eq:ch3}.
 }}
\end{center}
Note that Challenge 2 and the SDR probabilistic SINR constrained
problem \eqref{eq:sdr} are related via the following identification:
% Each probabilistic SINR constraint in \eqref{eq:sdr_SINR_con} can be rewritten in the form \eqref{eq:ch3}, where
\begin{subequations}\label{eq: definitions of qrs2}
\begin{align}
\displaystyle \Qb&= \frac{\epsilon_i^2}{3}~
\begin{bmatrix}
   {\rm Re}\left\{\frac{1}{\gamma_i}\Wb_i-\sum_{k\neq i}\Wb_k\right\} & -{\rm Im}\left\{\frac{1}{\gamma_i}\Wb_i-\sum_{k\neq
i}\Wb_k\right\} \\
   {\rm Im}\left\{\frac{1}{\gamma_i}\Wb_i-\sum_{k\neq
i}\Wb_k\right\} & {\rm Re}\left\{\frac{1}{\gamma_i}\Wb_i-\sum_{k\neq
i}\Wb_k\right\}
  \end{bmatrix}, \\
\rb  &= \frac{\epsilon_i}{\sqrt{3}}~ \begin{bmatrix}
   {\rm Re}\left\{\left(\frac{1}{\gamma_i}\Wb_i-\sum_{k\neq i}\Wb_k\right)\bar{\hb}_i\right\} \\
   {\rm Im}\left\{\left(\frac{1}{\gamma_i}\Wb_i-\sum_{k\neq
i}\Wb_k\right)\bar{\hb}_i\right\}
  \end{bmatrix},\\
  s&= \bar{\hb}_i^H\left(\frac{1}{\gamma_i}\Wb_i-\sum_{k\neq i}\Wb_k\right)\bar{\hb}_i
  -\sigma_i^2, ~~~ \rho = \rho_i.
\end{align}
\end{subequations}
To tackle Challenge 2 using the decomposition approach, we first observe that the
sum $\Psi = \eb^T\Qb\eb + 2\eb^T\rb$ can be written as
\begin{align*}
   \Psi =&~ \sum_{\ell=1}^n Q_{\ell\ell}e_{\ell}^2 + \sum_{1\le \ell\not=j\le n} Q_{\ell j}e_{\ell}e_j + 2\sum_{\ell=1}^n e_{\ell}r_{\ell} \\
   %\noalign{\medskip}
   =&~ \sigma_e^2 \sum_{\ell=1}^n Q_{\ell\ell} + \sum_{\ell=1}^n \left[ \left( \sum_{(j,j) \in \mathcal{A}_{\ell}} Q_{jj} (e_j - \sigma_e^2) \right) + \left( \sum_{(j,k) \in \mathcal{A}_{\ell}} Q_{jk}e_je_k \right) \right] + 2\sum_{\ell=1}^n e_{\ell}r_{\ell} \\
   %\noalign{\medskip}
   =&~ \sigma_e^2 \cdot \tr(\Qb) + \sum_{\ell=1}^n \Psi_{q\ell} + \Psi_l,
\end{align*}
where $\sigma_e^2 = {\mathbb E}\{ e_1^2 \}$ and the sets
$\mathcal{A}_1,\ldots,\mathcal{A}_{n}$ are defined as in Table
\ref{ColoringTable}.  In other words, if the $(j,k)$th entry of the
table is labeled $\mathcal{A}_{\ell}$, then $(j,k) \in \mathcal{A}_{\ell}$.
\begin{table}[t]
\begin{center}
\caption{Construction of the sets
$\mathcal{A}_1,\ldots,\mathcal{A}_n$.} \label{ColoringTable}
\begin{tabular}{|c||c|c|c|c|c|}
\hline
 & 1 &  2 & $\cdots$ & $n-1$ & $n$\\
\hline\hline
1 &  $\mathcal{A}_1$&  $\mathcal{A}_2$ &  $\cdots$ &  $\mathcal{A}_{n-1}$ & $\mathcal{A}_n$\\
\hline
2 & $\mathcal{A}_2$ &  $\mathcal{A}_3$ & $\cdots$ & $\mathcal{A}_n$ & $\mathcal{A}_{1}$\\
\hline
$\vdots$ & $\vdots$ & $\vdots$ & $\ddots$ & $\vdots$ & $\vdots$ \\
\hline
$n-1$ &$\mathcal{A}_{n-1}$&$\mathcal{A}_n$ & $\cdots$ &$\mathcal{A}_{n-3}$ & $\mathcal{A}_{n-2}$\\
\hline
$n$ &$\mathcal{A}_n$ & $\mathcal{A}_1$ & $\cdots$ &$\mathcal{A}_{n-2}$ & $\mathcal{A}_{n-1}$\\
\hline
\end{tabular}
\vspace{-0.4cm}
\end{center}
\end{table}

Using Table \ref{ColoringTable}, it is not hard to verify that each of the
terms $\Psi_{q1},\Psi_{q2},\ldots,\Psi_{qn},\Psi_l$ is a sum of
independent random variables.  Thus, by bounding their moment
generating functions and using an argument similar to that in the
previous subsection, we obtain the following method for Challenge 2
(again, see \cite{CSW11} for details):

~ \\
\noindent \fbox{\parbox{\linewidth}{ {\bf Method IV
for Challenge 2 (Decomposition into Independent Parts):} \ The
following feasibility problem is a convex restriction
of \eqref{eq:ch3} in Challenge 2:
\begin{align*}
 \text{Find}&~~ \Qb,\rb,s,\tb \\
 \text{s.t.}&~~
    s + \sigma_e^2 \cdot \tr(\Qb) \geq 2\sqrt{-\ln(\rho)} \cdot \sum_{\ell=0}^n
    t_{\ell}, \\%\notag
    &~~\sqrt{2} \|\rb\| \leq t_0,\\%\notag
    &~~\left( \sum_{(j,k)\in\mathcal{A}_{\ell}} v_{jk}^2Q_{jk}^2 \right)^{1/2}
    \leq t_{\ell},~\ell=1,\ldots,n,
\end{align*}
where $\sigma_e^2 = {\mathbb E}\{ e_1^2 \}$, $v_{jj}=1/\sqrt{8}$ and $v_{jk}=1$ if $j\not=k$, for
$(j,k)\in\mathcal{A}_{\ell}$ and $\ell=1,\ldots,n$.
}} ~

%%%%%%%%%%%%%%%%%%%%%%%%%%%%%%%%%%%%%%%%%%%%%%%%%%%%%%%%%%%%%%%%%%
\section{Simulation Results}\label{sec: simulation results}
%%%%%%%%%%%%%%%%%%%%%%%%%%%%%%%%%%%%%%%%%%%%%%%%%%%%%%%%%%%%%%%%%%

%In this section, we present some simulation results to examine the
%performance of the proposed RAR formulations.% The simulation setting
%is presented in the first subsection. Second subsection presents the
%performance comparison results of Method I to Method III under
%complex Gaussian CSI errors. The last subsection examines the
%performance of Method IV by assuming bounded, i.i.d. uniformly
%distributed CSI errors.

%%%%%%%%%%%%%%%%%%%%%%%%%%%%%%%%%%%%%%%%%%%%%%%%%%%%%%%%%%%%%%%%%%%%%%%%%%%%
%\subsection{Simulation Setting}
%%%%%%%%%%%%%%%%%%%%%%%%%%%%%%%%%%%%%%%%%%%%%%%%%%%%%%%%%%%%%%%%%%%%%%%%%%%%

%Simulations were carried out to evaluate the performance of the proposed RAR methods.
This section shows an extensive set of simulation results illustrating the performance of the proposed RAR methods.

Let us first describe the general simulation settings.
We employ a universal QoS specification for all users;
i.e., $\gamma_1=\cdots=\gamma_K \triangleq \gamma$,
$\rho_1=\cdots=\rho_K \triangleq \rho$.
The users' noise powers are identical and fixed at $\sigma_1^2=\cdots=\sigma_K^2=0.1$.
In each simulation trial,
the presumed channels $\{\bar\hb_i\}_{i=1}^K$ are randomly and independently generated according to the standard complex Gaussian distribution.

%In the simulations, the users are assumed to have the same noise
%power 0.1, i.e., $\sigma_1^2=\cdots=\sigma_K^2=0.1$ and request the
%same target SINR value, denoted by $\gamma\triangleq
%\gamma_1=\cdots=\gamma_K$. The user SINR outage probabilities are
%also set the same, i.e., $\rho\triangleq \rho_1=\cdots=\rho_K$. The
%coefficients of the preassumed channel $\{\bar\hb_i\}_{i=1}^K$ are
%generated according to the i.i.d. complex Gaussian distribution with
%zero mean and unit variance.

Next, we provide some implementation details of the RAR methods.
The RAR problems (those in Table I) are solved by the conic optimization solver $\texttt{SeDuMi}$ \cite{sedumi}, implemented through the now popularized and very convenient parser software $\texttt{CVX}$ \cite{cvx}.
%The popular convex optimization software \texttt{CVX} \cite{cvx}
%is used to solve the RAR problems (recall the problems in Table~\ref{table_SDRs}).
Then, we check whether a solution $(\Wb_1,\ldots,\Wb_K)$ to an RAR problem
is of rank one or not.
If yes, then the rank-one decomposition, $\Wb_i = \wb_i \wb_i^H$ $\forall i$,
is used to obtain a beamforming solution $(\wb_1,\ldots,\wb_K)$.
Otherwise, the Gaussian randomization procedure in Algorithm~\ref{alg1} is called to generate a feasible $(\wb_1,\ldots,\wb_K)$.
Numerically, we declare that $(\Wb_1,\ldots,\Wb_K)$ is of rank one if
the following conditions hold:
%\begin{align}
$$ \frac{\lambda_{\max}(\Wb_i)}{\tr(\Wb_i)}\geq 0.99 \quad\text{for all}~
i=1,\ldots,K;% \label{normalized maximum eigenvalue}
$$
%\end{align}
i.e., the largest eigenvalue of $\Wb_i$ is at least $100$ times
larger than any of the other eigenvalues.
Moreover, we say that an RAR method is feasible if
the RAR problem has a feasible solution
and the subsequent beamforming solution generation procedure
is able to output a feasible $(\wb_1,\ldots,\wb_K)$.

%For the proposed RAR formulations, we declare that the obtained
%solution $(\Wb_1,\ldots,\Wb_K)$ is of \emph{rank one} if the
%following condition is satisfied
%\begin{align}
%\frac{\lambda_{\max}(\Wb_i)}{\tr(\Wb_i)}\geq 0.99~ \text{for all}~
%i=1,\ldots,K, \label{normalized maximum eigenvalue}
%\end{align}
%which implies that the largest eigenvalue is around 100 times
%larger than any of the rest eigenvalues. By simulations, it is found
%that when this condition \eqref{normalized maximum eigenvalue} holds
%true, the beamforming vectors obtained by simple rank-one
%decomposition $\Wb_i = \wb_i \wb_i^H$ empirically satisfy the desired SINR
%satisfaction probabilities. Moreover, we say that a RAR formulation
%is \emph{feasible} for a channel realization $\{\bar\hb_i\}_{i=1}^K$ if
%the RAR formulation yields a rank-one solution. In the case that
%the RAR formulation yields a higher-rank solution, we say that it is
%feasible only if the Gaussian randomization procedure in Table
%\ref{Table specialized IPA} yields a rank-one feasible
%approximate solution.

The RAR methods are benchmarked against the probabilistic SOCP methods
in~\cite{Shenouda2008}.
The latter are also implemented by $\texttt{SeDuMi}$ through $\texttt{CVX}$.
%The latter are also implemented by \texttt{CVX}.
To provide a reference,
we also run a conventional perfect-CSI-based SINR constrained design
(e.g., \cite{WieselTSP06}), where the presumed channels
$\{\bar\hb_i\}_{i=1}^K$ are used as if they were perfect CSI.
We will call this the ``non-robust method'', for convenience.

%We will compare the performance of the proposed RAR formulations
%with the restrictive approximation method presented in
%\cite{Shenouda2008}, specifically, the formulation I in
%\cite{Shenouda2008}. The popular convex solver
%\texttt{CVX} \cite{cvx} is used to solve all the formulations under
%test.

%%%%%%%%%%%%%%%%%%%%%%%%%%%%%%%%%%%%%%%%%%%%%%%%%%%%%%%%%%%%%%%%%%%%%%%%%%%%
%\subsection{Performance Comparison Results for Complex Gaussian CSI
%errors}\label{simulation_complex_Gaussian}
%%%%%%%%%%%%%%%%%%%%%%%%%%%%%%%%%%%%%%%%%%%%%%%%%%%%%%%%%%%%%%%%%%%%%%%%%%%%
%
%\subsubsection{ Example 1}

\subsection{Simulation Example 1}

We start with the simple case of $N_t=K=3$; i.e., three antennae at the base station, and three users.
The CSI errors are spatially i.i.d. and have standard complex Gaussian distributions; i.e., $\Cb_1=\cdots=\Cb_K= \sigma_e^2 \Ib_{N_t}$, where $\sigma_e^2 > 0$ denotes the error variance.
We set $\sigma_e^2 = 0.002$.
The SINR requirement is $\gamma= 11$dB.
The outage probability requirement is set to $\rho= 0.1$, which is equivalent to having a $90\%$ or higher chance of satisfying the SINR requirements.

\begin{figure}[!t]
\begin{center}
    \resizebox{0.8\textwidth}{!}{
        %\psfrag{Nt=K}[Bc][Bc]{\Large $N_t=K$}
        %\psfrag{gamma (dB)}[Bc][Bc]{\Large $\gamma$ (dB)}
        \includegraphics{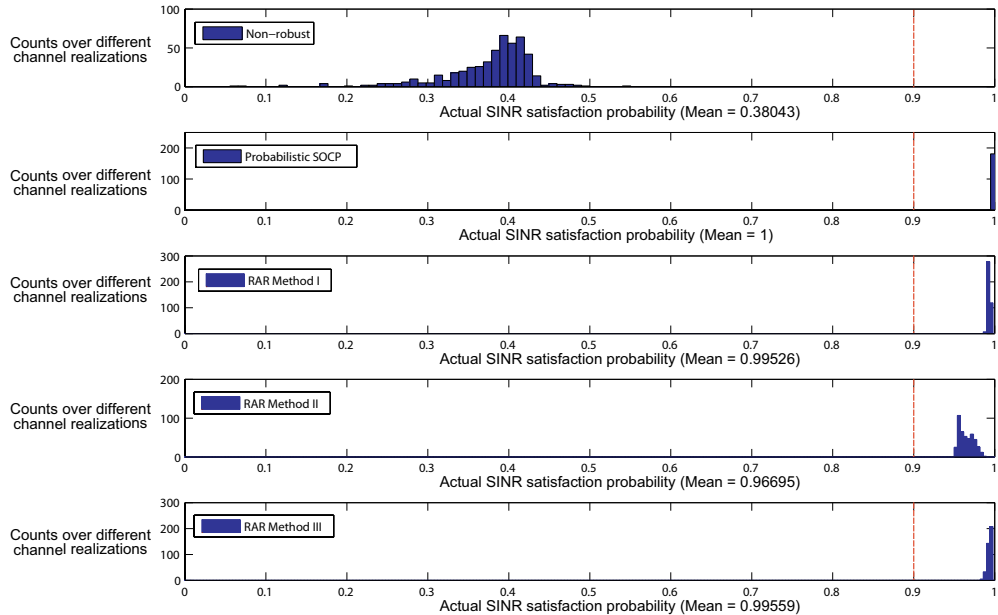}} \vspace{-0.5cm}
        \caption{Histograms of the actual SINR satisfaction probabilities of the various methods. $N_t=K=3$; $\Cb_1=\cdots=\Cb_K=0.002\Ib_{N_t}$; $\gamma=11$ dB; $\rho=0.1$.}
        \label{Empirical SINR distribution}\vspace{-0.6cm}
\end{center}
\end{figure}

First, we are interested in examining the actual SINR satisfaction probability,
${\rm Prob}\{ {\sf SINR}_i \geq \gamma \}$, of the various methods.
Figure~\ref{Empirical SINR distribution} shows the histograms of the actual SINR satisfaction probabilities over different channel realizations.
To obtain the histograms,
we generated $500$ realizations of the presumed channels $\{\bar\hb_i\}_{i=1}^K$.
Then, for each channel realization,
the actual SINR satisfaction probabilities of all methods
were numerically evaluated using $10,000$ randomly generated realizations of the CSI errors $\{\eb_i\}_{i=1}^{K}$,
which should be sufficient in terms of the probability evaluation accuracy.
Figure~\ref{Empirical SINR distribution} validates that
our RAR methods (and the existing probabilistic SOCP method) indeed
adhere to the $90\%$ SINR satisfaction specification.
There are two interesting observations, as can be seen from the figure.
The first is with the non-robust method.
While the non-robust method is, by nature, expected to violate the SINR outage specification,
its actual SINR satisfaction probabilities are below $50\%$ for most of the channel realizations,
which is severe.
This reveals that the perfect-CSI-based design can be quite sensitive to CSI errors.
The second is with the conservatism of the various robust methods.
The probabilistic SOCP method has its actual SINR satisfaction probabilities
concentrating at $100\%$, which indicates that it may be playing too safe
in meeting the outage specification.
By contrast, our RAR methods seem to be less conservative.
Particularly, among the three methods, RAR Method II (Bernstein-type inequality) appears to be the most relaxed as observed from its histogram.

%In the example, we start with a simple case
%of $N_t=K=3$, that is, three antennas in the base station, and three users. The covariance
%matrices of CSI errors $\{\eb_i\}_{i=1}^{K}$ are set to
%{\color{blue} $\Cb_1=\cdots=\Cb_K= 0.002\Ib_{N_t}$.
%(any justification for this number?)} The SINR outage probability
%$\rho$ is set to $0.1$, which means $90\%$ SINR satisfaction
%probability. We first examine whether the four methods
%under test can satisfy the SINR satisfaction probability 0.9. To do
%this, we generate 500 sets of channel realizations
%$\{\bar{\hb}_{i}\}_{i=1}^{K}$. For each channel realization, if the
%formulation under test is feasible, we further generate $10,000$
%sets of CSI errors $\{\eb_i\}_{i=1}^{K}$ following $\mathcal{CN}(
%\bm{0}, 0.002\Ib_{N_t})$ to evaluate the associated empirical SINR
%satisfaction probabilities. Figure \ref{Empirical SINR distribution}
%shows the histograms of the empirical SINR satisfaction
%probabilities of user 1, where the SINR requirement $\gamma$ is set
%to 11 dB. The result of the conventional non-robust design
%\cite{Farrokhi1998} is also shown in this figure.

%We can observe from Fig. \ref{Empirical SINR distribution} that the
%non-robust design cannot meet the required $0.9$ SINR satisfaction
%probability, and there are more than 50\% chance that SINR outage occurs.
%By contrast, the
%four robust formulations can always achieve over $0.9$ SINR
%satisfaction probabilities for the users. This simulation result
%demonstrates the importance of the robust designs in providing
%guaranteed receiver performance.

Next, we investigate the conservatism of the various robust methods
by evaluating their feasibility rates;
i.e., the chance of getting a feasible beamforming solution under different channel realizations.
Similar to the last investigation,
$500$ channel realizations were used.
The obtained result is shown in
Figure~\ref{Gaussian CSI case}(a),
where the feasibility rates of the various methods are plotted against the SINR requirements $\gamma$.
Remarkably, the three RAR methods yield feasibility rates much higher than that of the probabilistic SOCP method.
In particular, RAR Method II has the best feasibility rate performance,
which is consistent with
%its more relaxed SINR satisfaction probabilities noticed
the SINR satisfaction probability result we noted
in Figure~\ref{Empirical SINR distribution}.
The feasibility rates of RAR Methods I and III are a close match:
For $\gamma > 9$dB, RAR Method I slightly outperforms RAR Method III;
for $\gamma \leq 9$dB, we see the converse.

\begin{figure}[t!]
\begin{center}
{\subfigure[][]{\resizebox{0.45\textwidth}{!}{\includegraphics{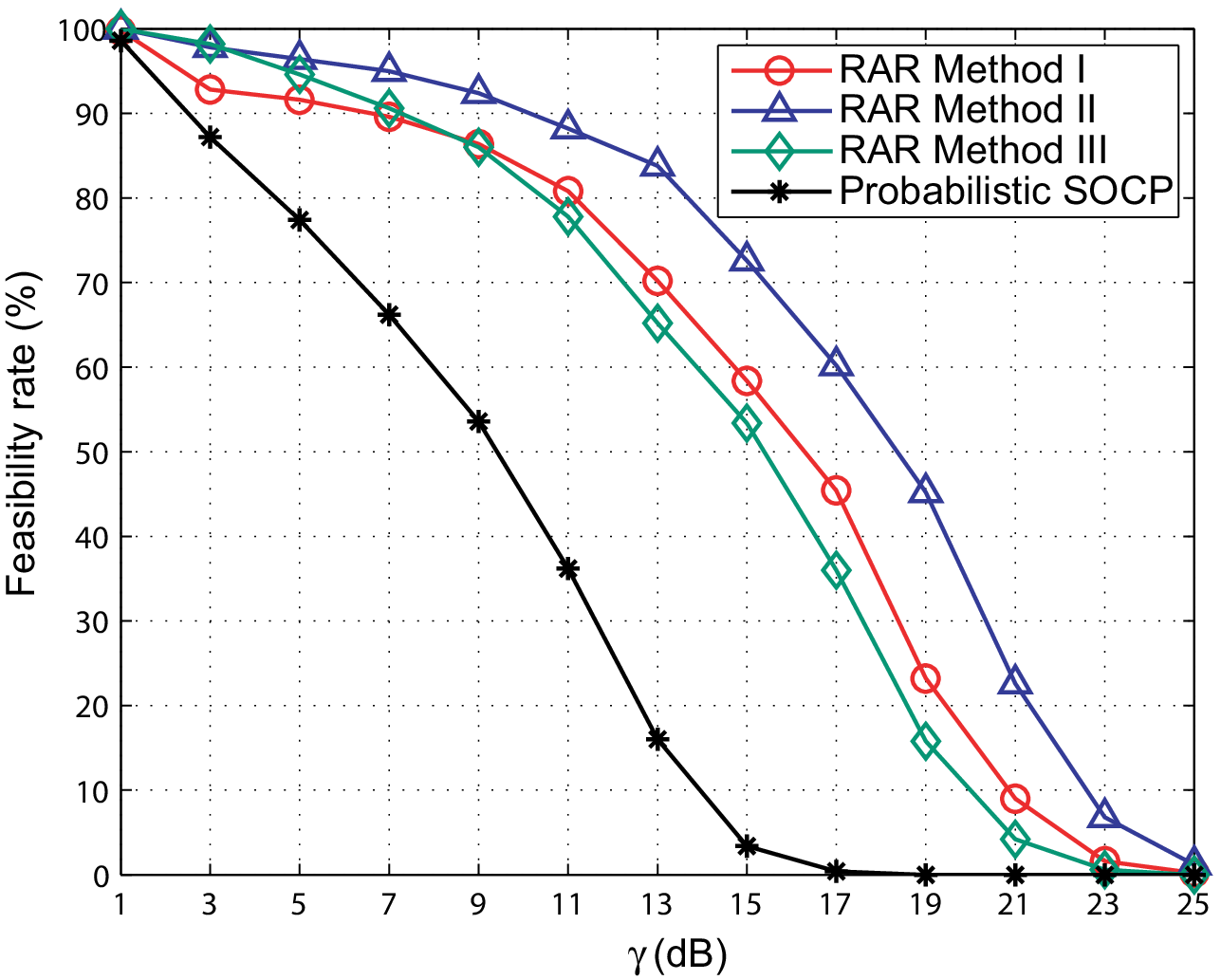}}}}
    \hspace{+0.2cm}
{\subfigure[][]{\resizebox{0.45\textwidth}{!}{\includegraphics{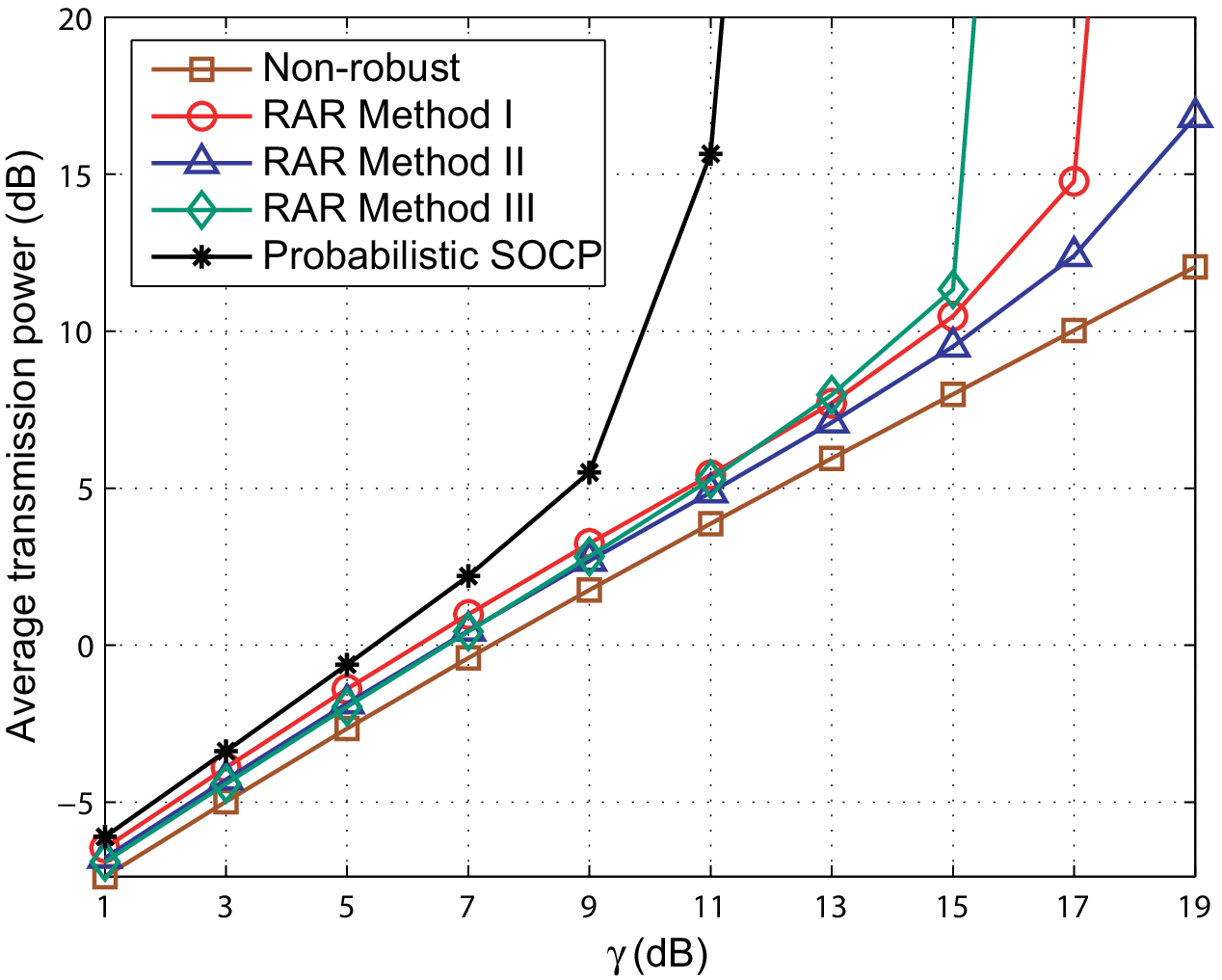}}}}
\vspace{+0.2cm}
\end{center}\vspace{-0.8cm}
\caption{Feasibility and transmit power performance of the various methods.
$N_t=K=3$; $\rho=0.1$; spatially i.i.d. Gaussian CSI errors with $\sigma_e^2= 0.002$.
%\gamma=11$dB;
%$\Cb_1=\ldots=\Cb_K=0.002\Ib_{N_t}$;
}
\vspace{-0.6cm} \label{Gaussian CSI
case}
\end{figure}

In addition to the feasibility rate, it is important to examine the transmit power consumptions of the design solutions offered by the various robust methods.
Figure~\ref{Gaussian CSI case}(b) shows the result.
It was obtained based on channel realizations for which
all methods yield feasible solutions at $\gamma= 11$dB; $181$ such realizations were found out of $500$ realizations (the same realizations used in the last result in Figure~\ref{Gaussian CSI case}(a)).
As can be seen from Figure~\ref{Gaussian CSI case}(b),
RAR Method II yields the best average transmit power performance,
followed by RAR Methods I and III (with Method I exhibiting noticeably better performance for $\gamma > 15$dB),
and then the probabilistic SOCP method in~\cite{Shenouda2008}.
As a reference, we also plot the transmit powers of the non-robust method
in the figure, so as to get an idea of how much additional transmit power
would be needed for the robust methods to accommodate the outage specification.
We see that for $\gamma \leq 11$dB, the transmit power difference between an RAR method and the non-robust method is about $1.5$dB,
which is reasonable especially when compared to the probabilistic SOCP method.
The gaps gradually widen, otherwise. %as $\gamma$ increases.
This seems to indicate that imperfect CSI effects are more difficult to cope with when we demand higher SINRs.

\begin{figure}[!t]
\begin{center}
    \resizebox{0.5\textwidth}{!}{
        \psfrag{Nt=K}[Bc][Bc]{\Large $N_t=K$}
        %\psfrag{gamma (dB)}[Bc][Bc]{\Large $\gamma$ (dB)}
        \includegraphics{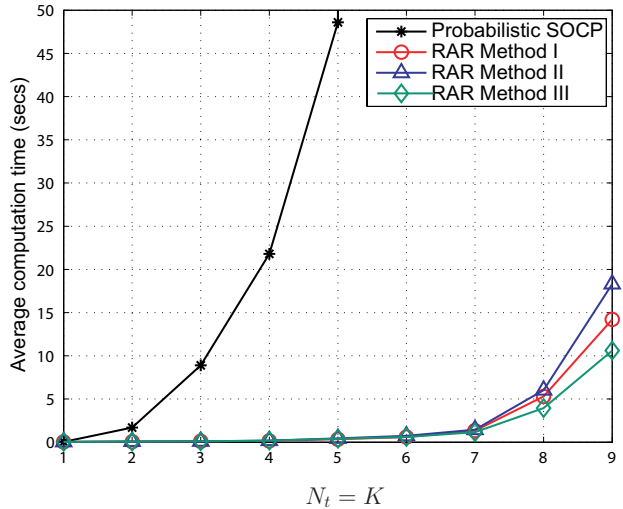}} \vspace{-0.2cm}
        \caption{Average runtimes of the various methods. %$\Cb_1=\ldots=\Cb_K=0.002\Ib_{N_t}$; $\gamma=7$dB; $\rho=0.1$.
        }\label{Complexity}\vspace{-0.5cm}
\end{center}
\end{figure}

%From the performance results above,
%we have the following performance ranking of the robust methods in general:
%RAR Method II, RAR Method I, RAR Method III, and probabilistic SOCP.
Now, let us consider the computation times of the various robust methods.
%Figure~\ref{Complexity} gives the runtimes with respect to the problem size $N_t =K$.
The result is illustrated in Figure~\ref{Complexity}.
To obtain this result, we use a desktop PC with $2.13$GHz CPU and $3$GB RAM. Moreover, instead of calling the convenient parser $\texttt{CVX}$, we use direct $\texttt{SeDuMi}$ implementations of all the methods, done by careful manual problem transformation and programming. The reason of doing so is to bypass parsing overheads, which may result in unfair runtime comparisons.
%The computational times were measured on a laptop PC with $1.9$GHz CPU and $4$Gb RAM.
From the figure, we see that the runtime ranking,
from the shortest to longest, is:
RAR Method III, RAR Method I, RAR Method II, and the probabilistic SOCP method.
Interestingly and coincidently, the runtime ranking of the RAR methods is exactly the opposite of their performance ranking we see in the previous simulation result.

As the last result in this example,
we numerically inspect a technical issue that has much implication to the RAR approach---
how frequent do the RAR problems yield rank-one solutions.
Recall that rank-one RAR solution instances have the benefits
that the beamforming solution generation is simple
(simple rank-one decomposition, no Gaussian randomization),
and that feasibility of the RAR problem directly implies that of beamforming solution generation.
Table~\ref{Table: Gaussian distributed CSI error} shows the result.
There is a ratio in each field.  The denominator is the realizations count
for which the RAR problem is feasible,
and the numerator is the realizations count for which the RAR problem yields a rank-one solution.
Again, $500$ channel realizations were used.
Curiously, almost all the fields in Table~\ref{Table: Gaussian distributed CSI error} indicate rank-one solution all the time.
We encountered only one non-rank-one instance out of $480$ for the setting of $\rho=0.01$, $\gamma=3$dB, RAR Method II.
We therefore conclude, on the basis of numerical evidence,
that occurrence of high-rank RAR solutions is very rare
for the unicast outage-based SINR constrained problem considered here.

\begin{table}[t]
  \caption{Ratios of rank-one RAR solutions.
  %$N_t=K=3$; $\Cb_1=\ldots=\Cb_K=0.002\Ib_{N_t}$.
  }
    \begin{center}
    \begin{tabular}{|m{2cm}||m{1cm}|m{1cm}|m{1cm}|m{1cm}||m{1cm}|m{1cm}|m{1cm}|m{1cm}|}
    \hline
    ~~~~~~~~$\rho$ & \multicolumn{3}{c}{~~~~~~~~~~~~~0.1} & & \multicolumn{3}{c}{~~~~~~~~~~~~~0.01} & \\
    \hline
    ~~~~~$\gamma$ (dB) & ~~~~3 & ~~~~7 & ~~~11 & ~~~15 & ~~~~3 & ~~~~7 & ~~~11 & ~~~15 \\
    \hline
    ~~{\bf Method I} & 464/464 & 448/448 & 404/404 & 292/292 & 450/450 & 424/424 & 343/343 & 225/225 \\
    \hline
    ~~{\bf Method II} & 489/489 & 475/475 & 441/441 & 363/363 & {\color{red} 479/480} & 463/463 & 428/428 & 322/322 \\
    \hline
    ~~{\bf Method III} & 488/488 & 453/453 & 389/389 & 267/267 & 476/476 & 421/421 & 306/306 & 144/144 \\
    \hline
    \end{tabular}
    \end{center}
\label{Table: Gaussian distributed CSI error}\vspace{-0.5cm}
\end{table}

%Since the proposed RAR formulations use SDR, it is interesting to check how often the proposed RAR formulations can yield rank-one solutions. Table \ref{Table: Gaussian distributed CSI error} shows
%the simulation results. In particular, each
%element in the tables is a ratio of the number of channel
%realizations for which a rank-one solution is obtained to the
%number of tested feasible channel realizations. As seen from Table \ref{Table: Gaussian distributed CSI error}, for
%the case of $\rho=0.1$, all the three RAR formulations yield
%rank-one solutions for the tested problem instances. Interestingly,
%for the case of $\rho=0.01$, we can see that rank-one solutions are
%still obtained in most of the cases, except for
%$\gamma=3$dB where there is a channel realization for which Method II
%yields a higher-rank solution. We can see from this table that it is very rare to obtain higher-rank solutions

\subsection{Simulation Example 2}
This example considers more challenging settings, described as follows:
$N_t = K = 8$;
spatially correlated CSI errors where $\Cb_1=\cdots=\Cb_{K}=\Cb_e$,
%\begin{equation}
$$ [ \Cb_e ]_{m,n} = \sigma_e^2 \times 0.9^{|m-n|}, $$
%\end{equation}
and $\sigma_e^2 = 0.002$;
$\rho= 0.01$ (or $99\%$ SINR satisfaction probability).
We do not run the probabilistic SOCP method in~\cite{Shenouda2008}, since, as seen in Figure~\ref{Complexity},
it is computationally very demanding for large problem sizes.
The same simulation method in Simulation Example 1 was used to produce the results here.
Figure~\ref{Gaussian CSI error difficult} shows the resulting feasible rates and average transmit powers.
A minor simulation aspect with the transmit power performance plot in Figure~\ref{Gaussian CSI error difficult}(b) is that we choose $\gamma= 7$dB as the pick-up point of feasible channel realizations of all the methods.
We can see that, once again, RAR Method II offers superior performance over the others.
Another observation is that RAR Method III manages to outperform RAR Method I this time.

Figure~\ref{Gaussian CSI error difficult2} illustrates another set of results,
where we increase the CSI error variance $\sigma_e^2$ from $0.002$ to $0.01$.
The number of users is set to $K= 6$.
The feasible realizations pick-up point is $\gamma= 13$dB.
%The SINR point  is chosen to pick feasible channel realizations for producing the average transmit power plot.
We can see similar performance trends as in the previous result in Figure~\ref{Gaussian CSI error difficult}.

\begin{figure}[!t]
\begin{center}
    {\subfigure[][]{\resizebox{0.45\textwidth}{!}{\includegraphics{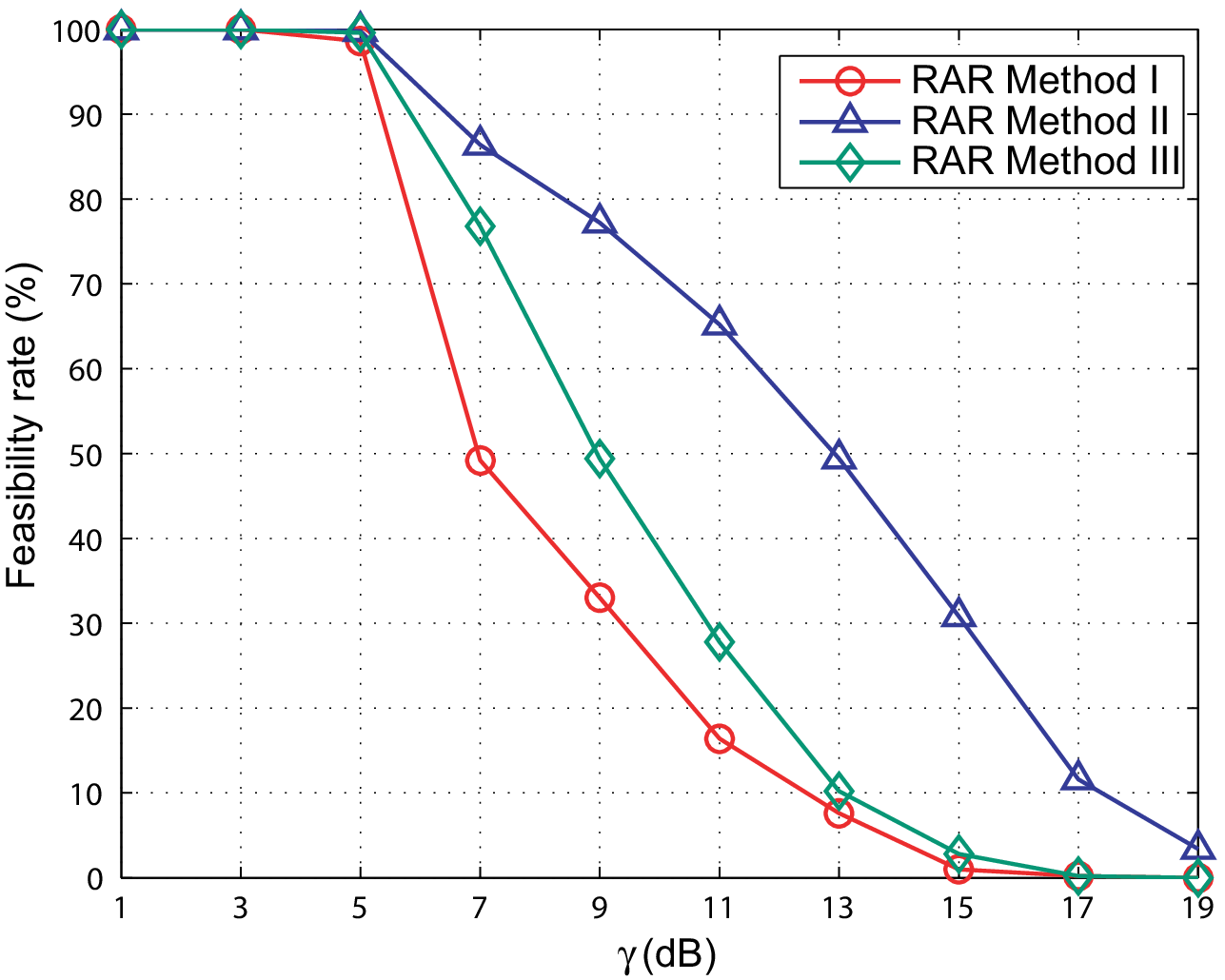}}}}
    \hspace{+0.2cm}
    {\subfigure[][]{\resizebox{0.45\textwidth}{!}{\includegraphics{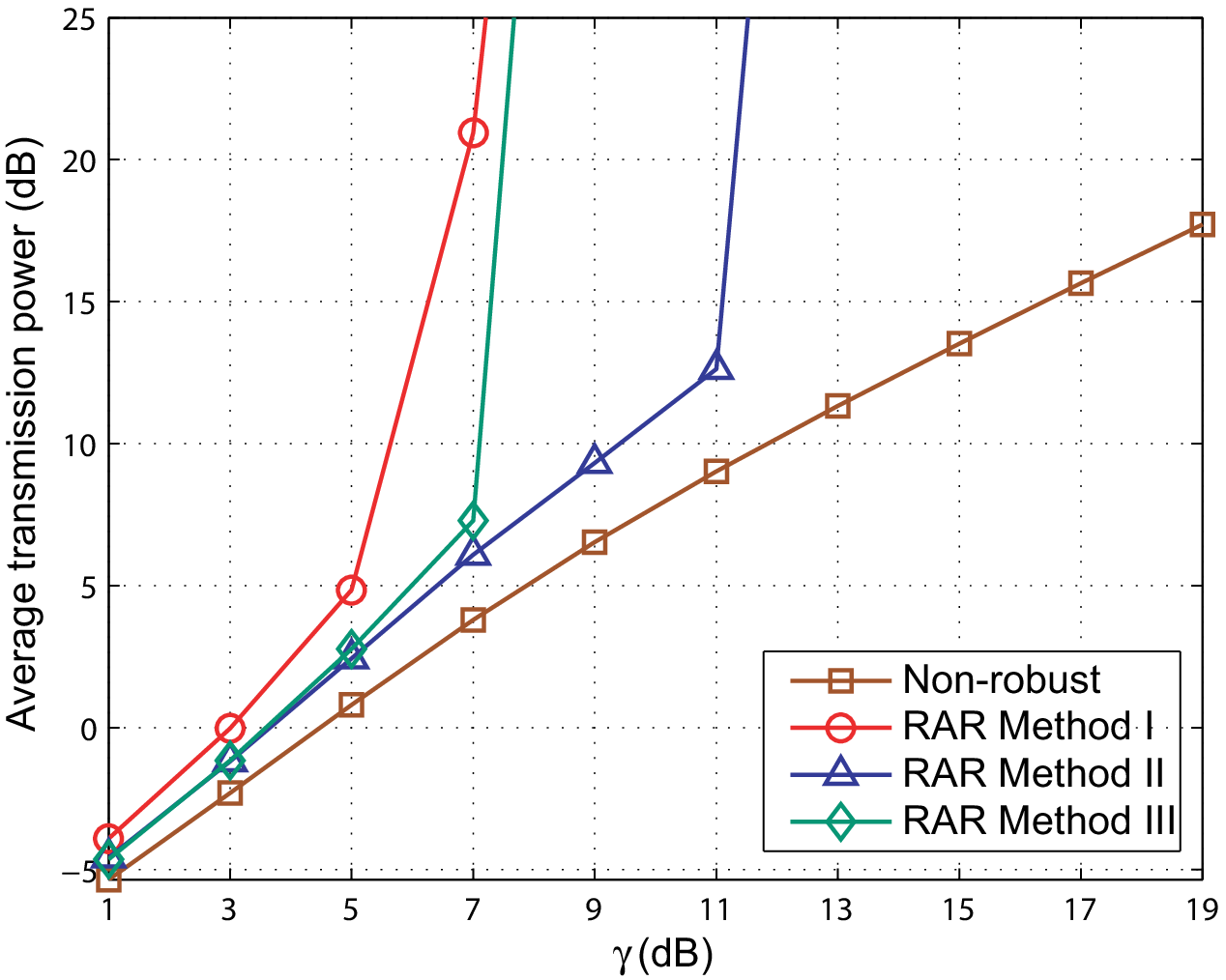}}}}
\end{center}\vspace{-0.4cm}
\caption{Performance under spatially correlated Gaussian CSI errors. $N_t=K=8$; $\rho=0.01$; $\sigma_e^2= 0.002$.
}\vspace{-0.5cm} \label{Gaussian
CSI error difficult}
\end{figure}

\begin{figure}[!t]
\begin{center}
    {\subfigure[][]{\resizebox{0.45\textwidth}{!}{\includegraphics{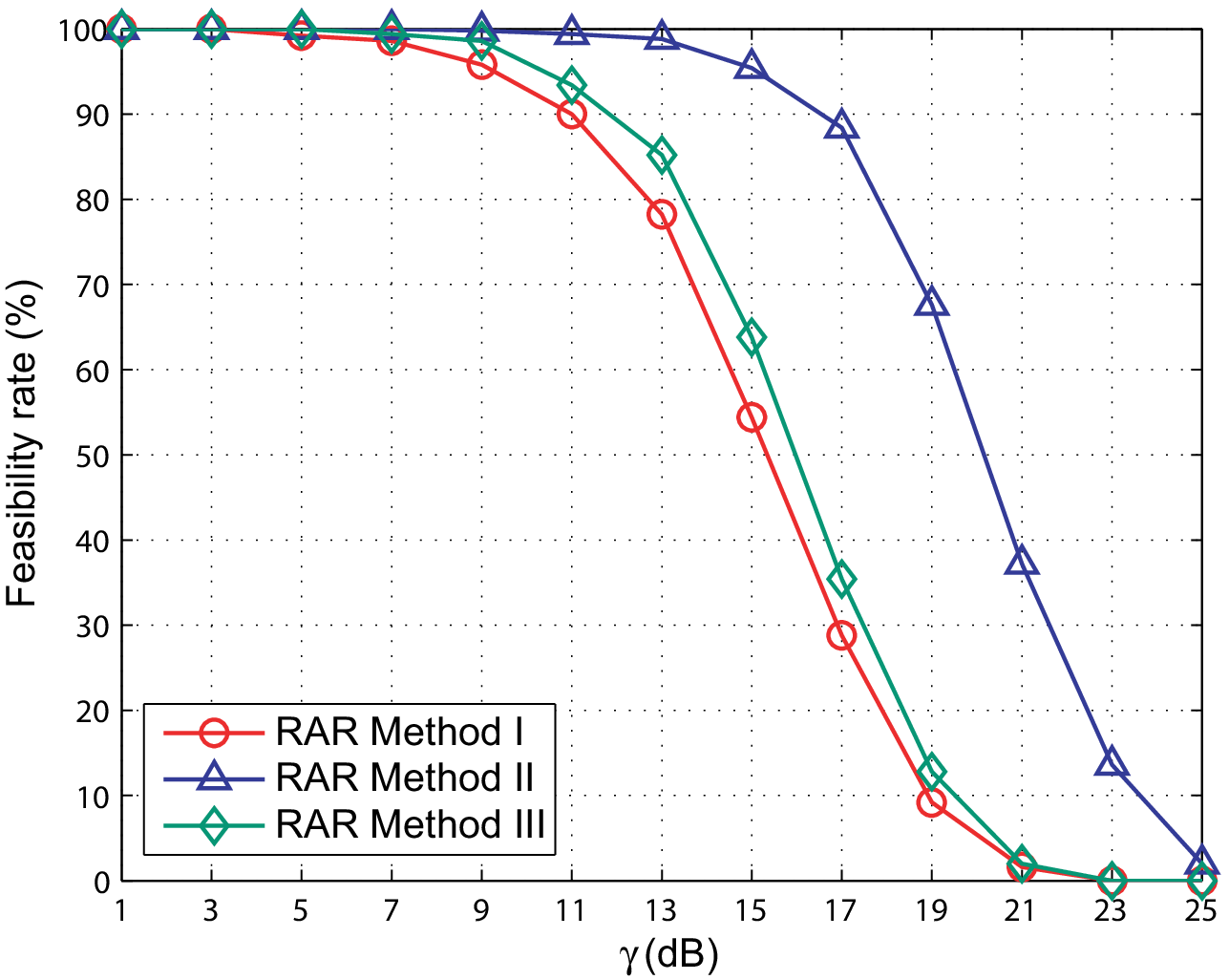}}}}
    \hspace{+0.2cm}
    {\subfigure[][]{\resizebox{0.45\textwidth}{!}{\includegraphics{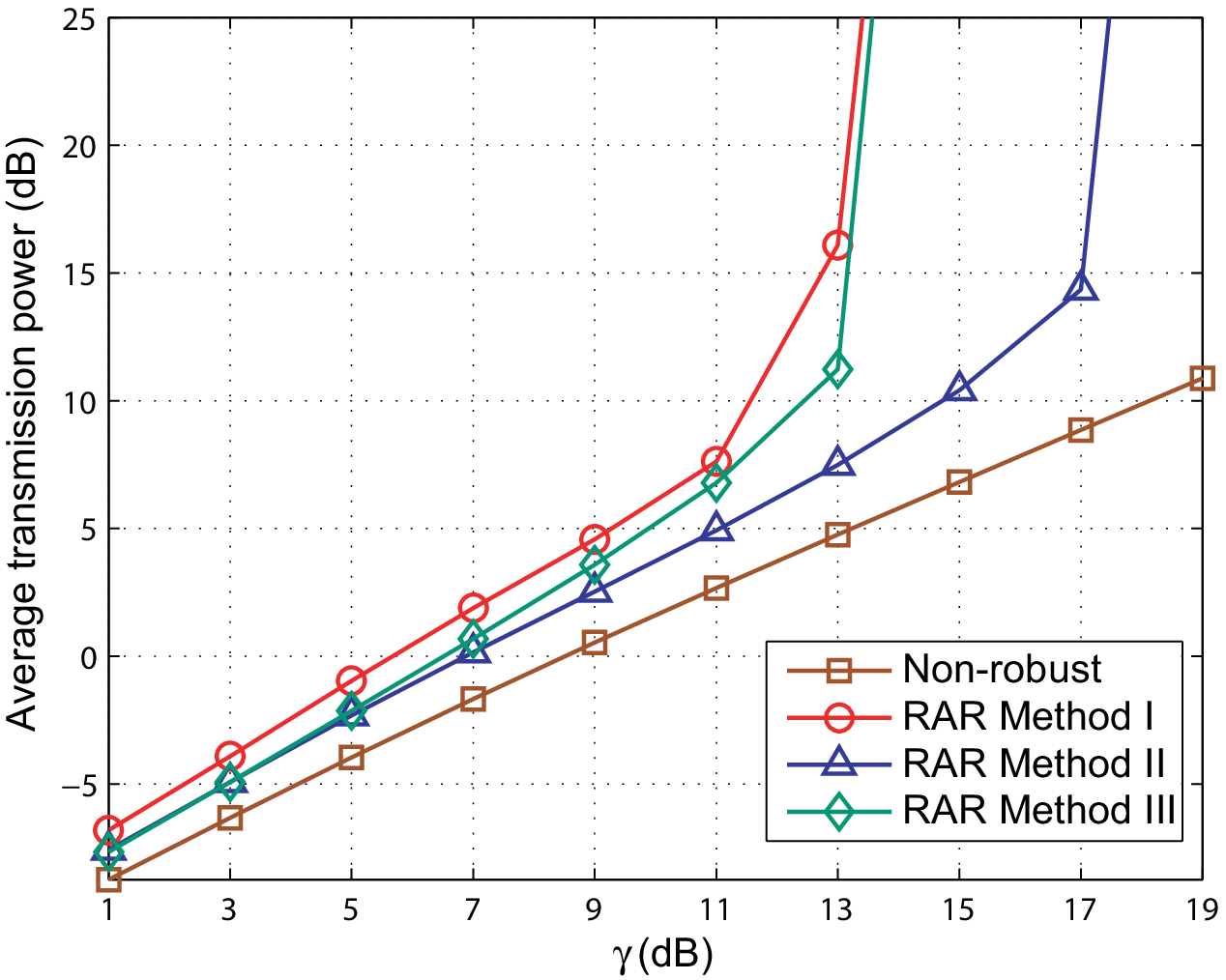}}}}
\end{center}\vspace{-0.4cm}
\caption{Performance under spatially correlated Gaussian CSI errors.
$N_t=8$; $K=6$; $\rho=0.01$; $\sigma_e^2= 0.01$.} \label{Gaussian
CSI error difficult2}\vspace{-0.5cm}
\end{figure}

\subsection{Simulation Example 3}

One might have noticed from Simulation Example 1, Figure~\ref{Empirical SINR
distribution},
that none of the robust methods yield actual SINR satisfaction probabilities at $1-\rho$.
This means that the robust methods are, to certain extent, conservative.
If more computations are allowed, this conservatism may be mitigated by running a bisection scheme.  Such a scheme was first proposed in~\cite{Ben-tal2009} in the context of chance constrained optimization, and then was adopted in~\cite{Shenouda2008} for the probabilistic SOCP method.
The idea is to fine tune some design parameters relevant to the outage requirement.
If a design solution is found to satisfy the outage specification well,
then we adjust the design parameters to relax the outage requirement
(e.g., for RAR Method I, decreasing $d_1,\ldots,d_K$)
and rerun the design problem.  Otherwise, we do the opposite and rerun
the design problem.
The above step is done repeatedly following a bisection search,
requiring the design problem to be solved multiple times.
The bisection search also requires a validation procedure
for satisfiability of the outage specification,
which can be done using the Monte-Carlo based validation procedure in \cite{Ben-tal2009}.
It is clear from the above discussion that the bisection scheme can also be
applied to all the RAR methods.
For more complete descriptions of the bisection scheme in the context of the probabilistic SINR constrained beamforming problem,  readers are referred to \cite{Shenouda2008,Wang2010,Wang2011}.

%A key ingredient is a Monte Carlo-based statistical validation procedure \cite{Ben-tal2009}, which is used to provide a (empirical) certificate on the satisfactions of the outage-based SINR constraints.
%If the certificate is yes, then we relax the outage requirement through altering the relevant system parameter (e.g., for RAR Method I, decreasing $d_1,\ldots,d_K$)
%and rerun the design problem.
%If the certificate is no, then we tighten the outage requirement by doing the opposite.
%This parameter tuning is done iteratively,
%requiring solving the design problems multiple (and many) times.
%Apparently, this idea can be adopted to all the RAR methods.
%Readers are referred to \cite{Shenouda2008,Wang2010,Wang2011} for more complete descriptions.

\begin{figure}[!t]
\begin{center}
    %\psfrag{gamma (dB)}[Bc][Bc]{\Large $\gamma$ (dB)}
    %\psfrag{Feasibility rate (\%)}[Bc][Bc]{\Large Feasibility rate (\%)}
{\subfigure[][]{\resizebox{0.45\textwidth}{!}{\includegraphics{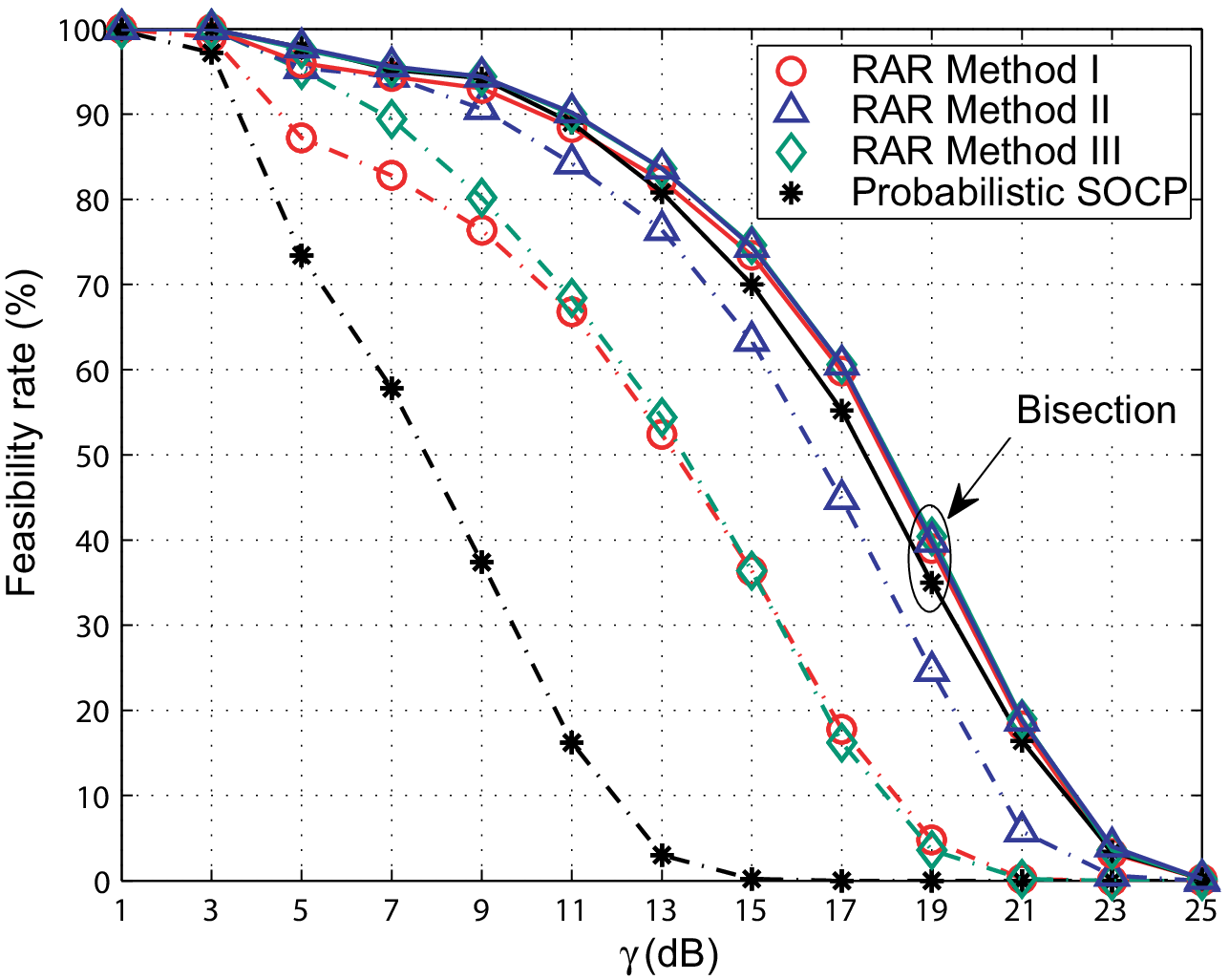}}}}
    \hspace{+0.2cm}
    %\psfrag{CSI error power}[Bc][Bc]{\Large $\sigma_e^2$}
    %\psfrag{[2]}[Bc][Bc]{\Large \cite{Shenouda2008}}
    %\psfrag{Feasibility rate (\%)}[Bc][Bc]{\Large Feasibility rate (\%)}
{\subfigure[][]{\resizebox{0.45\textwidth}{!}{\includegraphics{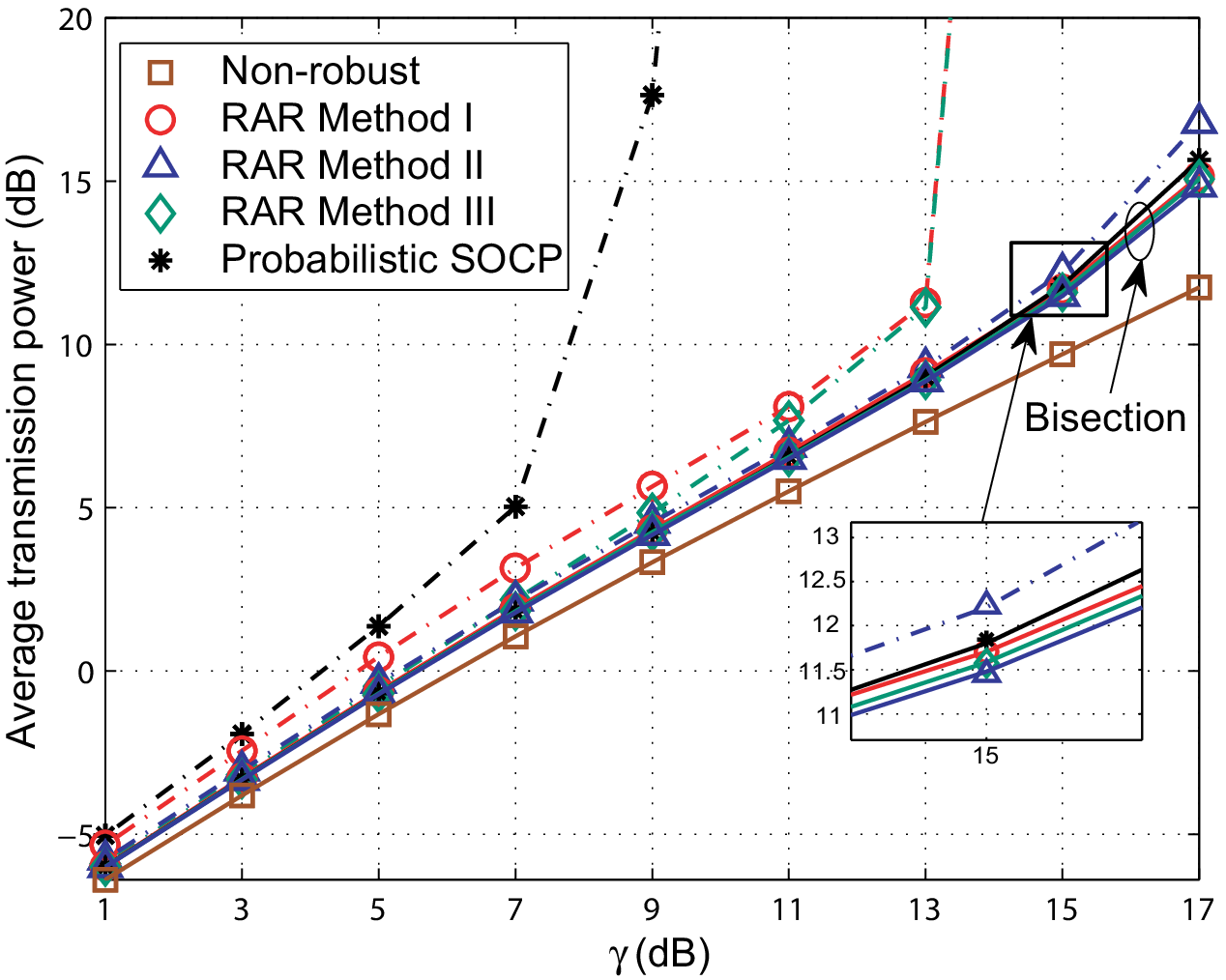}}}}
    %\vspace{0.5cm}}
    %\psfrag{gamma (dB)}[Bc][Bc]{\Large $\gamma$ (dB)}
    %\psfrag{[2]}[Bc][Bc]{\Large \cite{Shenouda2008}}
    %\psfrag{Feasibility rate (\%)}[Bc][Bc]{\Large Feasibility rate (\%)}
\end{center}\vspace{-0.4cm}
\caption{Performance with bisection. $N_t=K=5$; $\rho=0.1$; %$\Cb_1=\ldots=\Cb_K=0.002\Ib_{N_t}$.
spatially i.i.d. Gaussian CSI errors with $\sigma_e^2= 0.002$.
}\label{Bisection}\vspace{-0.5cm}
\end{figure}

Figure~\ref{Bisection} shows how bisection may improve the performance.
The simulation settings are $N_t=K=5$, $\rho=0.1$,
$\Cb_1=\cdots=\Cb_K=0.002\Ib_{N_t}$, and
the feasible realizations pick-up point at $\gamma= 9$dB (for the transmit power performance evaluations only).
We can see that all the robust methods, after applying the bisection scheme, exhibit improved performance.
Notwithstanding, we also see that RAR Method II without bisection already gives performance quite on a par with the bisection-aided methods.

\subsection{Simulation Example 4}
This example demonstrates the performance of RAR Method IV,
which handles elementwise i.i.d. bounded CSI errors with unknown distribution.
We test the method using elementwise i.i.d. uniform CSI errors,
where the real and imaginary parts of all $\eb_i$ are independent and uniformly distributed on $[-\epsilon, \epsilon]$ with $\epsilon > 0$.
The probabilistic SOCP method in~\cite{Shenouda2008} also has a version for i.i.d. uniform CSI errors and is included in our simulation.

The simulation settings are: $N_t=K=3$, $\rho=0.1$, $\epsilon=0.02$,
and the feasible realizations pick-up point at $\gamma= 7$dB.
The result, presented in Figure~\ref{Uniform_CSI error}, illustrates that
RAR Method IV provides much better performance than the probabilistic SOCP method.
Table~\ref{Table: Uniformly distributed CSI error} shows the ratios of getting a rank-one RAR solution,
where we see clearly that encountering high rank RAR solutions is rare.

%%%%%%%%%%%%%%%%%%%%%%%%%%%%%%%%%%%%%%%%%%%%%%%%%%%%%%%%%%%%%%%%%%%%%%%%%%%%%%%%%
%\subsection{Performance Comparison Results for Uniformly Distributed CSI Errors}
%%%%%%%%%%%%%%%%%%%%%%%%%%%%%%%%%%%%%%%%%%%%%%%%%%%%%%%%%%%%%%%%%%%%%%%%%%%%%%%%%
%
%In this section, we examine the performance of the proposed Method
%IV in the presence of i.i.d. bounded CSI errors. In particular,
%we assume that the real and imaginary parts of each of the CSI
%errors follow the i.i.d. uniform distribution in the support
%$[-\epsilon,\epsilon]$ where $\epsilon>0$. Since Formulation I in
%\cite{Shenouda2008} can also handle this CSI error, we compare it with the proposed Method
%IV.
%
%Figure \ref{Uniform_CSI error} presents the simulation results for
%$N_t=K=3$, $\rho=0.1$ and $\epsilon=0.02$, by testing over $500$
%sets of channel realizations $\{\bar{\hb}_i\}_{i=1}^{K}$. The
%results in Fig. \ref{Uniform_CSI error}(b) is obtained by averaging
%over 168 sets of feasible channel realizations at $\gamma=7$ dB. It
%can be seen from this figure that the proposed Method IV
%significantly outperforms the Formulation I in \cite{Shenouda2008}
%in terms of both feasibility rate and average transmission power. As
%observed from Table \ref{Table: Uniformly distributed CSI error}, on
%the other hand, the proposed Method
%IV can yield rank-one solutions in most of the cases.%

\begin{figure}[!t]
\begin{center}
{\subfigure[][]
{\resizebox{0.45\textwidth}{!}{\includegraphics{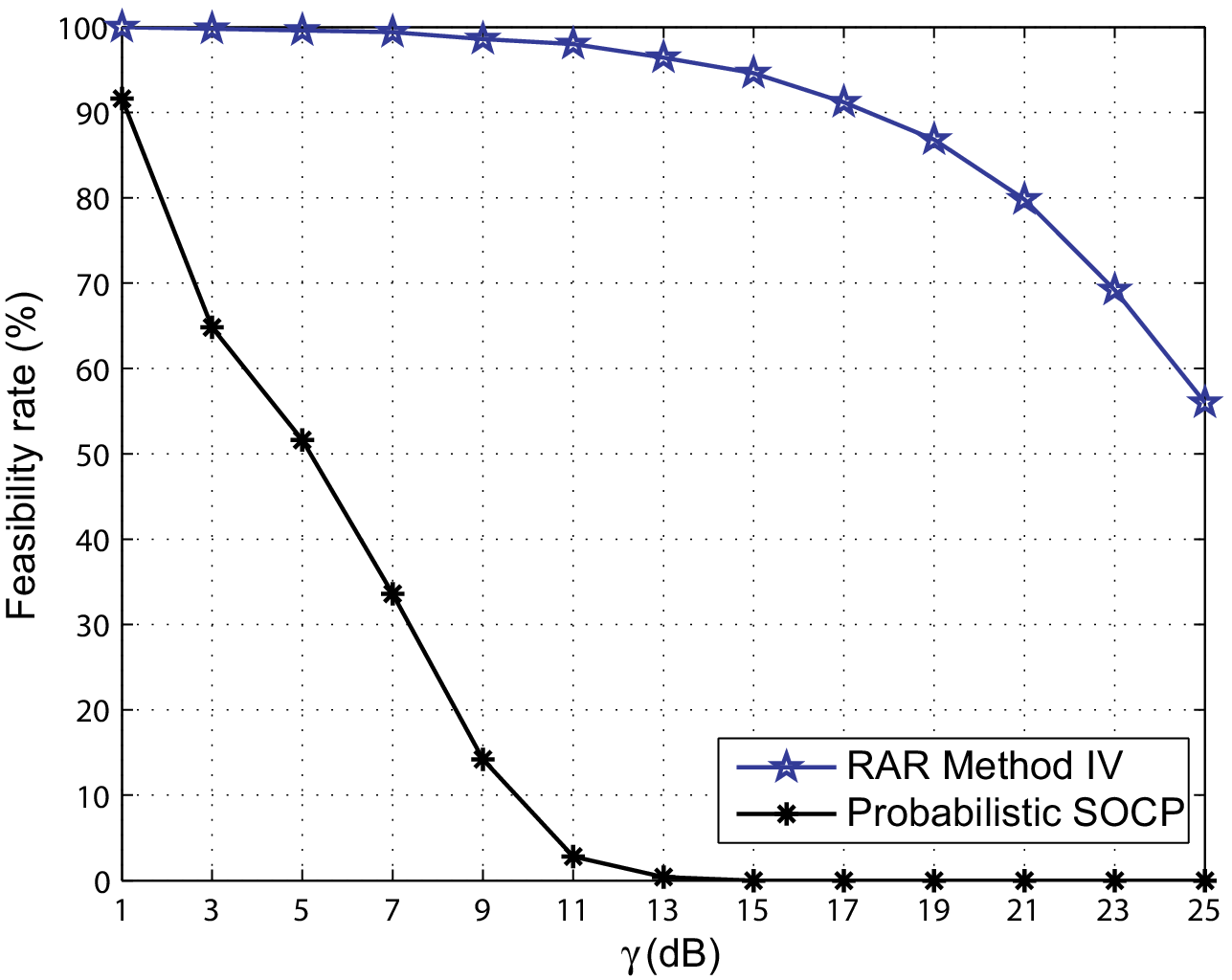}}}}
    \hspace{+0.2cm}
%{\subfigure[][ $\gamma=5$ dB]
%{\resizebox{0.45\textwidth}{!}{\includegraphics{./Figures/Fig6c}}}}
%    \hspace{+0.2cm}
{\subfigure[][]{\resizebox{0.45\textwidth}{!}{\includegraphics{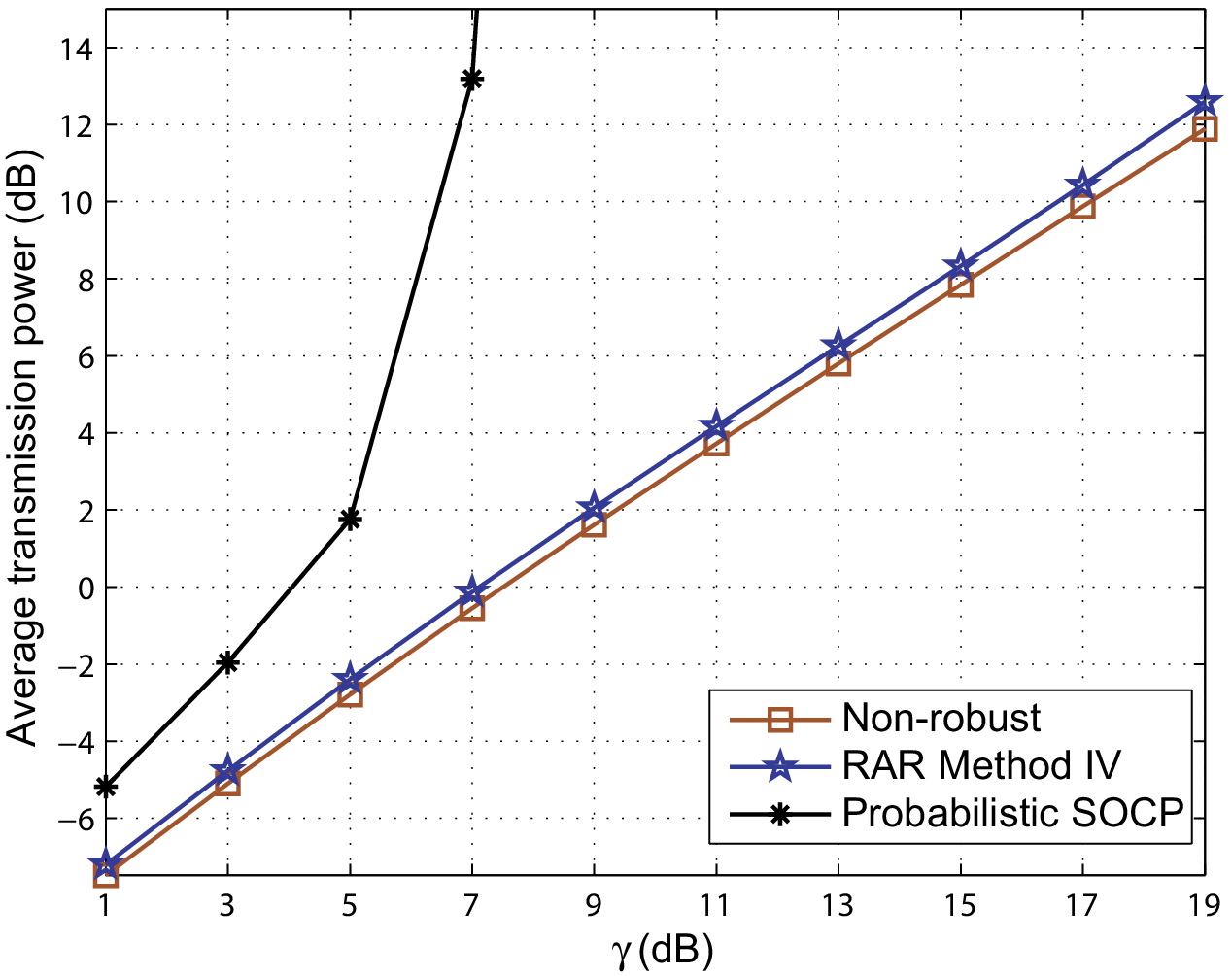}}}
    \vspace{0.5cm}}
%{\subfigure[][$\gamma=5$ dB]
%{\resizebox{0.45\textwidth}{!}{\includegraphics{./Figures/Fig6d}}}}
%\vspace{+0.2cm}
\end{center}
\caption{Performance under i.i.d. uniform CSI errors. $N_t=K=3$; $\rho=0.1$; $\epsilon=0.02$.}\vspace{-0.5cm} \label{Uniform_CSI
error}
\end{figure}

\begin{table}[t]
  \caption{Ratio of rank-one RAR solutions.
  %$N_t=K=3$; $\epsilon=0.02$; $\rho=0.1$.
  }
    \begin{center}
    \begin{tabular}{|m{2cm}||m{1cm}|m{1cm}|m{1cm}|m{1cm}|m{1cm}|m{1cm}|m{1cm}|m{1cm}|}
    \hline
    ~~~~~$\gamma$ (dB) & ~~~~1 & ~~~~3 & ~~~~5 & ~~~~7 & ~~~~9 & ~~~11 & ~~~13 & ~~~15 \\
    \hline
    ~~{\bf Method IV} & 500/500 & {\color{red} 498/499} & 498/498 & 497/497 & 493/493 & 490/490 & 482/482 & 473/473 \\
    \hline
    \end{tabular}
    \end{center}
\label{Table: Uniformly distributed CSI error}\vspace{-0.5cm}
\end{table}

\section{Conclusions}\label{sec: conclusions}
Motivated by the presence of CSI errors in practical systems and the need to avoid substantial SINR outages among users, we studied a probabilistic SINR constrained formulation of the transmit beamforming design problem.  Although such formulation can safeguard each user's SINR requirement, it is difficult to process computationally due to the SINR outage probability constraints.  To circumvent this, we proposed a novel relaxation-restriction (RAR) approach, which features the use of semidefinite relaxation techniques, as well as analytic tools from probability theory, to produce efficiently computable convex approximations of the aforementioned probabilistic formulation.  One of our main contributions is the development of three methods--- namely, sphere bounding, Bernstein-type inequality, and decomposition--- for processing the probabilistic SINR constraints.  Our simulation results indicated that the proposed RAR methods provide good approximations to the probabilistic SINR constrained problem, and they significantly improved upon existing methods, both in terms of solution quality and computational complexity.

At the core of our technical development is a set of tools for constructing efficiently computable convex restrictions of chance constraints with quadratic uncertainties.  An interesting future direction would be to apply these new tools to other transmit beamforming formulations, such as those arising from the frontier cognitive radio and multicell scenarios, or perhaps even other signal processing applications.  %This would be an interesting future direction.

\bibliography{ref_PCRBF}

\end{document}